\documentclass{elsart}
\usepackage{graphicx}
\usepackage{rotating}
\usepackage{amssymb}

\newcommand{\be}{\begin{equation}}
\newcommand{\ee}{\end{equation}}

\begin{document}
\begin{frontmatter}

\title{Organization of networks  with tagged nodes and  biased links: {\it a priori} distinct communities.
 \\ The case of Intelligent Design Proponents and Darwinian Evolution Defenders }

\author{G. Rotundo$^{1,a}$, M. Ausloos$^{2,3,b}$} 
\address{
$^{1}$Faculty of Economics, University of Tuscia, via del Pa\-ra\-di\-so
47,\\ I-01100 Viterbo, Italy
\\ $^{2}${\it now at:}  7 rue des Chartreux, B-4122 Plainevaux, Belgium 
\\ $^{3}${\it  previously at: } GRAPES@SUPRATECS, Universit\'e de Li\`ege,  B5a Sart-Tilman, B-4000 Li\`ege, Euroland
}
\ead{$^{a}$giulia.rotundo@uniroma1.it;$^{b}$marcel.ausloos@ulg.ac.be }




\date{\today}
\maketitle
\vskip 0.5truecm

\begin{abstract}
Among  topics of opinion formation it is of interest to observe
the  characteristics   of  networks with {\it a priori} distinct communities.  As an illustration, we report on the citation network(s) unfolded in the recent decades 
through web available  works belonging to selected members of
  the   Neocreationist and Intelligent Design  Proponents (IDP)  and the Darwinian
  Evolution Defenders (DED) communities.
  An  adjacency matrix  of {\it tagged nodes} is first constructed; it is not symmetric. A generalization of considerations pertaining to the  case of networks with {\it biased links}, directed or undirected,  is  thus presented.

  The main characteristic coefficients describing   the structure   of such  partially directed networks with tagged nodes are outlined. The structural features are discussed searching for statistical aspects,  equivalence or not of subnetworks through the degree distributions, each network assortativity, the global and local clustering coefficients and the Average Overlap Indices. The   various closed and open triangles made from nodes, moreover distinguishing the community, are especially listed to calculate  the clustering characteristics.  The distribution of elements in the  $rectangular$ submatrices are specially examined since they represent inter-community connexions. The emphasis being on distinguishing the number of vertices belonging to a given community.  Using such informations one can distinguish between opinion leaders,  followers and main rivals and briefly interpret  their relationships through psychological-like conditions intrinsic to behavior rules in either community.  Considerations on other controversy cases  with similar social constraints are outlined, as well as suggestions on further, more general, work deduced from our observations on such networks.

\end{abstract}


\end{frontmatter}%

\section{Introduction  }\label{sec:intro}
In modern statistical physics nowadays \cite{stauffer1}, networks \cite{pastor} underlying
opinion formation of agents located at nodes \cite{1JSM}, with  links defined from data pertaining to
econophysics \cite{econophysics}
and/or
sociophysics \cite{sociophysics} have gathered much interest.
There are many applications in line \cite{costa}. This has led to
a flurry of theoretical works, most of them assuming a very large population of
interacting agents. 
All these studies concern the organizational processes of
populations, on networks or lattices
\cite{APE98,pnas.99.02.7821hub,PNAS-Netw-04slanina}. However, not all agent based systems need to be studied on large scale-free
networks, as if looking for some thermodynamic limit. On the contrary,  finite size networks  with agent small connectivity values are known to be more realistic \cite{milgram,killworth}.

Networks are usually composed of a large number of
internal components, i.e. nodes and links, which can be used to describe a wide
variety of systems of high  intellectual  and technological
importance. Relevant questions pertain to the dynamics of collective
properties, not only of agents on the network \cite{majoritymodel}, but also for the network
structure itself \cite{lambiotte1,lambiotte2}.  Recent results on the dynamics of social
networks   \cite{froncszak} suggest  the occurrence of  either
discontinuous, continuous, and high order phase transitions and coexistence
phase states in a large class of models   \cite{vicsek}. This
is also  similar to features found in percolation and  nucleation-growth problems.

Characteristics of small world networks (SWN) \cite{SWN},
as introduced by Watts and Strogatz 
might be   close to  describing social reality. It has been  argued,
though with more self-conviction than proof, that SWN are useful
to describe population opinion switches \cite{svenson02}. Indeed
opinion formation through $small$ communities can be discussed on
small networks \cite{1JSM} and still retain much value. Cases
of interest  pertain to the adherence, e.g.,     to religious denominations (or sects)
\cite{religion1}, ideology struggles \cite{dv00},   language spreading \cite{steele2} or  policy analysis \cite{PNAS99.02.7267_moss}, as far studied off-network, like in a mean field approximation.

 Up to now one has mainly searched for communities on networks  inside  rather restrictedly defined characteristics of node and/or link sets.  However nodes could have  several {\it a priori} tags.  A generalization of the  concept of networks to the case for example in which nodes are Ising-like spins has been already considered \cite{svenson02}, - but rather from a strictly magnetic point of view, without  obvious connexion to any societal case. Yet it is obvious that networks made of nodes with tags are very numerous, considering the status or the opinion of an individual, more generally speaking $agent$.
Many   examples of agents having even more than a binary set of {\it degrees of freedom}  exist in on a network. The intensity of each tag can be also much varied, as in wealth, religious or language tags.
Moreover links between tagged nodes of different natures  might also occur.

  Whence it is easily admitted that networks with {\it a priori} distinct communities  are frequent \cite{portercommunitystructurePhA386,communitystructurePorter0902,implicitcommIJMPC}.
For example,  a society can   be roughly considered to be composed of males and females.  Other cases exist. All our  readers belong to  networks made of friends and enemies. A network can also be made of  persons speaking one orseveral languages: restricting the present examples to mainly   bilingual populations, or countries;  see cases   like Belgium, Ireland or Canada, - all having  even marked differences. Networks of citizens belonging to  one or another party ideology, like republicans and democrats, or leftists and rightists, are also common.
A case of members with drastically different opinion   is that of
  the   Neocreationist and Intelligent Design  Proponents (IDP)  and the Darwinian
  Evolution Defenders (DED).

    It  is   of  general interest to observe the evolution of topical subjects pertaining to the nature of how science is perceived or understood, either by scientists or by the public
at large \cite{ballcriticalmass}, and why/how such opinions prevail and others disappear. One such topics 
is $creationism$ indeed.
It is opposed to aspects like Darwinian Evolution (DE). In recent years, creationism has been rejuvenated into a concept called Intelligent Design (ID), such that it pretends to be some kind of scientific alternative to DE and give some sort of interpretation to the big bang, and its consequences.
This
{\it per se} raises surely scientific (and other) questions \cite{reftomedia}.   The historical perspective is reduced here below to the main
aspects pertinent to  our present considerations:  for comprehensive completeness  those are only presented in Appendix A following \cite{garcia}.

No need to get further involved here in the pro and con, though one could, - being motivated by the affair, both as scientists and members of a monotheist culture. Several fundamental reasons why the subject is controversial, to say the least, have been often  tackled in a mediatic way,
hardly through unbiased scientific research. Letters (to the editors), papers, media appearance \cite{hurd,miller} by true scientists or politicians \cite{bosetti} or others are numerous on the subject for the last 15 years or so, enough to provide data to be analyzed within modern lines, as in statistical mechanics and scientometrics.
Indeed much work in statistical physics attempts nowadays to reconcile intuitive or qualitative features, sometimes stylized facts, with simple models still driving into complexity aspects.

 This DE-ID controversy subject of intense interest is thereby considered  here below, from the point of view of two small world networks  \cite{SWN,svenson02},  making a larger one. 
 Basically there are two communities, loosely connected, but with members  hardly evolving in opinion but strongly arguing against the opposite one held by the other rival community.
  The   following study is carried out by analyzing the structural   properties of the citation network unfolded in the recent decades (1990-2007) by web available
 works belonging to members of  IDP and DED groups.

  The data set acquisition and its limits are recalled in Sec. \ref{sec:dataset}. We discuss 2 communities (or 2 opinions)
 existing on 2 subnetworks with approximately the same sizes (37 for  IDP and 40  for DED). 
The methodology is based on constructing an  adjacency matrix and examining the distribution of $(i,j)$ elements in submatrices.
We use notations like   ''directed links'' (DL), i.e.  those which obvioulsly have a direction from $j$ to $i$
and ''undirected links'' (UL)  when   $both$  ($i,j$)  and
($j,i$) linkages exist.
Since the links may be, we emphasize, directional, but we do not consider their intensity nor frequency (thus there is no weight) nor timing  of the citations (thus we do not consider time lags), we refer to these networks as made of  merely $biased$ links.

It seems rather appropriate  to publish   the whole adjacency matrix, following   the data gathering methodology
presented in \cite{garcia}; see  Appendix B.
It might be expected to become as useful as the karate club  data \cite{karate} (which had
44 nodes) or the the acquaintance network of Mormons (with 43 nodes) \cite{mormons}.
 The network construction
follows in lines with studies on large-scale networks, like
co-authorship networks \cite{newman,iina2}. 

 The structural aspects are next discussed searching for statistical aspects,  equivalence or not of subnetworks, in Sec. \ref{sec:stat}.  First  it is searched whether it   can be unambiguously proved  
 that  the degree distribution(s) or the link distribution(s) follow  simple theoretical laws.

 As a matter of fact, understanding some of the characteristics of the community made of the ID proponents and DE defenders  requires accounting for their mutual interaction. As far as we know this aspect of the so called controversy has remained largely unexplored,  except in \cite{garcia}, 
 but appears of general scientific value,  whatever the pro and con arguments on the intellectual ideas in the communities.  Thus the present work aims to contribute to introducing a quantitative approach to the analysis  of the interaction between two biased groups, in  small  networks.
Let it be here pointed out that
 triplets of nodes are   particularly examined in order to emphasize the inter- and intra-community connexions   through one
assortativity  coefficient \cite{assortativity,mixing}, two clustering coefficients \cite{SWN} and one overlap index  \cite{gligorausloos}. In order to do so, one has to generalize the nomenclature of triplets, whence of triangles,  formed by nodes  which can belong to two distinct communities and    tied by UL or DL. 
The notation nomenclature  convention aspect is shown in Fig. 8.

 Practically for such a study, i.e. when considering   networks with distinct communities, it is necessary to generalize the usual characteristic coefficients of networks describing either their structure or some dynamics to the present case of so called networks with tagged nodes and biased links.
 Using such informations one can distinguish between leaders, rivals  and followers. Moreover having in mind some {\it presumably accepted behavioral rules} of both communities, one can briefly interpret the relationships through the usual ethics and adequate psychological conditions of the members of either group.


In Sec. \ref{sec:conclusions}, some further discussion on the statistical mechanics of this illustrative case is presented in line with general considerations.
Since it is intuitively obvious that there is no phase transition to be expected, in the sense of \cite{1JSM}  because the members of such communities are pretty much behaving as in a very deep potential well, one has  not to elaborate here  on  the dynamics of opinion formation. There is neither much change nor fluctuations in the node state, whatever the link weights or number.  However it can be imagined that the opinions can evolve if considerations as outlined in \cite{galamdebates} are taken into practice.  They are reformulated for the present context in the conclusion section. 

 No need to say that most of the ideas so below developed can encompass many (rigid) networks, not only   the DE-ID controversy, but also others   formed by scientists with {\it a priori} constraints on rival scientific work. Moreover the present study suggests to examine  more general cases in which the nodes have many tags, with different intensity ranges, whence systems which belong to  universal classes other than the symmetric  1/2-Ising spin. 

\section{The data set }\label{sec:dataset}

 First the main
actors of the ID movement were selected through the ID web pages and
their corresponding links  so far helped by an article
  by  
Pennock \cite{Pennock}.
Next,  a search of citations   was  made through the  Scholar Google search engine.
 Directed and undirected interactions between agents are
  found through citations of the other (or one's own)
ideas, as expressed through various media. 
We selected 37 IDP and 40 DED \cite{garcia}. The examined time range goes from  Oct. 01 till
Nov 15, 2007.

In so doing, an adjacency matrix
$M=(m_{ij})\in R^{77 \times 77}$ is built, see Appendix B. 
The matrix elements $m_{ij}$ take the value 1 or 0 depending on
whether or not  a citation of $i$ by $j$  has taken place.
Therefore, the number of
links going into a node $i$, i.e. the $in-degree$, represents the number of  authors
that cite $i$. The number of links exiting from a node $j$, i.e. the $out-degree$,
represents the number of authors that $j$ cites.  First we accepted self-citations, but they can be disregarded, see below.
 Recall that we consider   ''directed links'' (DL) and ''undirected links'' (UL).  In brief,
the existence of a DL from $j$ to $i$ - beyond meaning that $j$ cites $i$, implies that $j$ is $never$ cited by $i$.
 The number of citations might be asymmetric, but here no weight is given to a link nor to a $direction$: the number of citations and their timing are here taken as irrelevant.  The resulting graphs are thus of the binary directed nonweighted network types.  
The interconnectivity between different communities presented as encounter frequency,  describing the probability a person encountering a stranger
in another  community has been discussed in \cite{implicitcommIJMPC}. However in the present case the encounter is very biased. Thus a discussion and variant of \cite{implicitcommIJMPC} is left for further work.

Obviously the matrix $M$ is square but not  symmetric. For ease of the data
analysis, we have gathered in the low   ranks of   $M$   the IDP (rows
and columns  from 1 till 37), and the  DED in the upper ranks (rows
and columns  from 38 till 77). Each  agent    has received an arbitrary index in its subset. By writing

\begin{equation} M=\left( \begin{tabular}{ll} $C$ & $A$  \\  $B$ & $D$ \end{tabular}\right)
\end{equation}\label{M}

 we evidence the two possible communities or subgraphs described  respectively by matrices $C$ (= $creationists$) and $D$ (= $Darwinians$), i.e.
  $C \, \in \, R^{37\times 37} $
  contains all the intra-citations among the IDP community;
 $D \, \in \, R^{40\times 40} $ contains the intra-citations among the DED  community, but
   $A$ and $B$ are rectangular matrices describing intercommunity  links.

For further discussion,
let us also introduce the   matrix   $M_0$ such that all diagonal terms
are $0$, i.e. not considering any self-citation, i.e. we
define

\begin{equation} \label{M0}
M_0=\left( \begin{tabular}{ll} $ C_0$  &   $A$   \\  $ B $ & $D_0$
\end{tabular} \right)
\end{equation}

It is also of interest to define  a matrix emphasizing the links between communities, i.e. the matrix
\begin{equation} \label{F}
F=\left( \begin{tabular}{ll} 0 & $A$  \\  $B$ & $0$
\end{tabular} \right).
\end{equation}

\begin{figure}
\includegraphics[height=16cm,width=16cm]{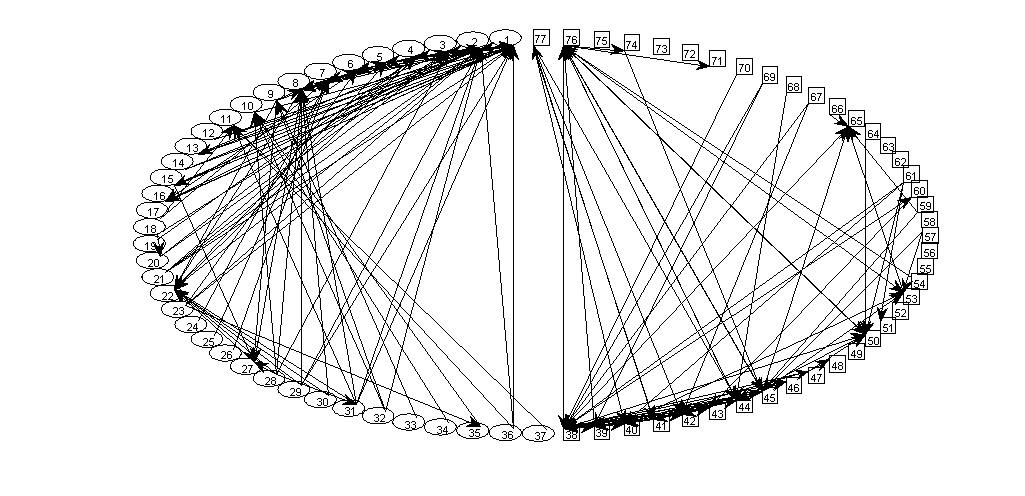}
\caption{  Intra-community links: (lhs) network of 37 agents (circles) corresponding to the IDP  community, i.e. the $C_0$  matrix; (rhs) network of 40 agents (squares) corresponding to the DED community, i.e.the $D_0$ matrix. In each community there are several   nodes/agents not linked to its community: 5, 11, 12, 14, 17, 18, 19, 21, 23,  24, 25, 26, 33, 34, 37, and 48, 49, 52, 54, 55, 56, 57, 59, 61, 62, 63, 64, 67, 68, 69, 70, 72, 73, 75 respectively. An arrow starting from $ j$ and pointing to $i$ is drawn if $m_{ij}=1$ and $m_{ji}=0$, for so called directed links (DL) } \label{fig:C0D0}
\end{figure}

\begin{figure}
\includegraphics[height=16cm,width=16cm]{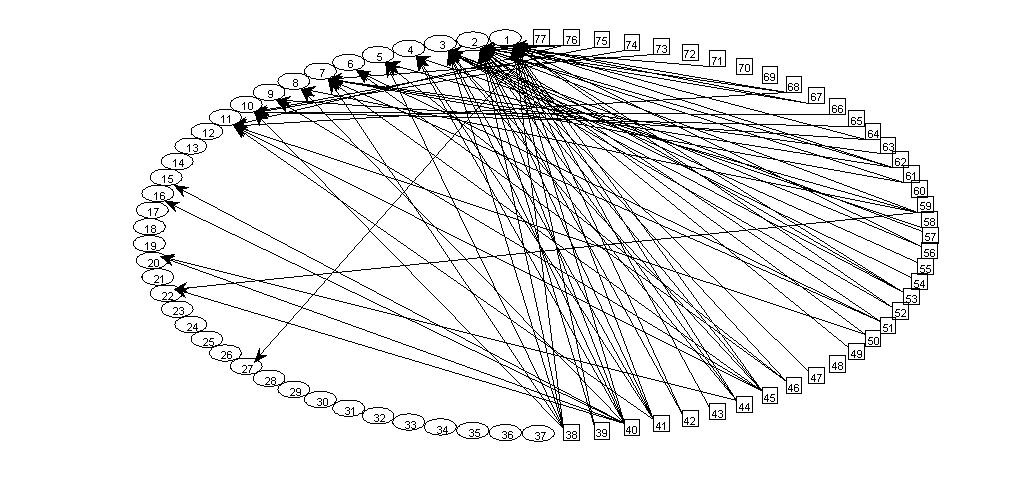}
\caption{Network  corresponding to off-diagonal block  rectangular matrix A, i.e. when  some DED cites some IDP.  An arrow indicates a  directed link (DL)}
\label{fig:A}
\end{figure}

\begin{figure}
\includegraphics[height=16cm,width=16cm]{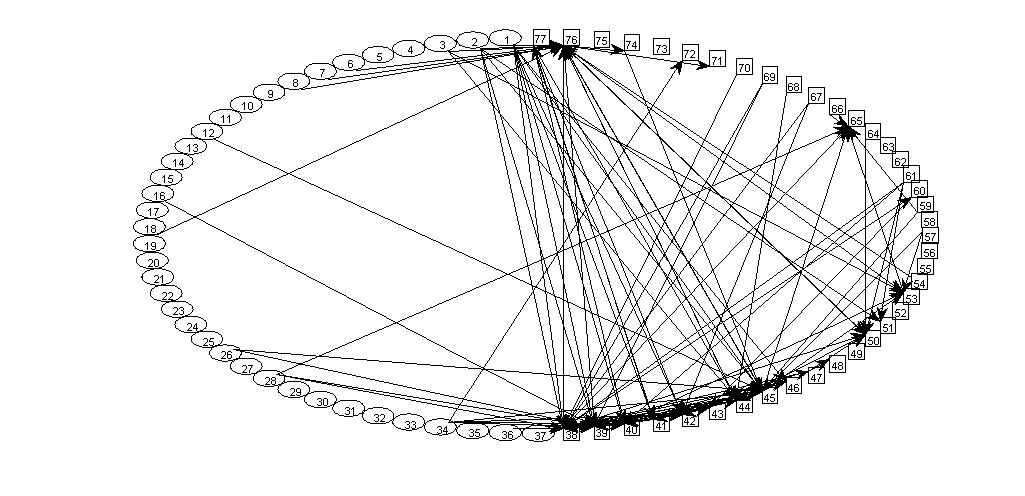}
\caption{Network corresponding to off-diagonal block  rectangular matrix B, i.e. when  some IDP cites some DED.  An arrow indicates a  directed link (DL)}
\label{fig:B}
\end{figure}

\begin{figure}
\includegraphics[height=16cm,width=16cm]{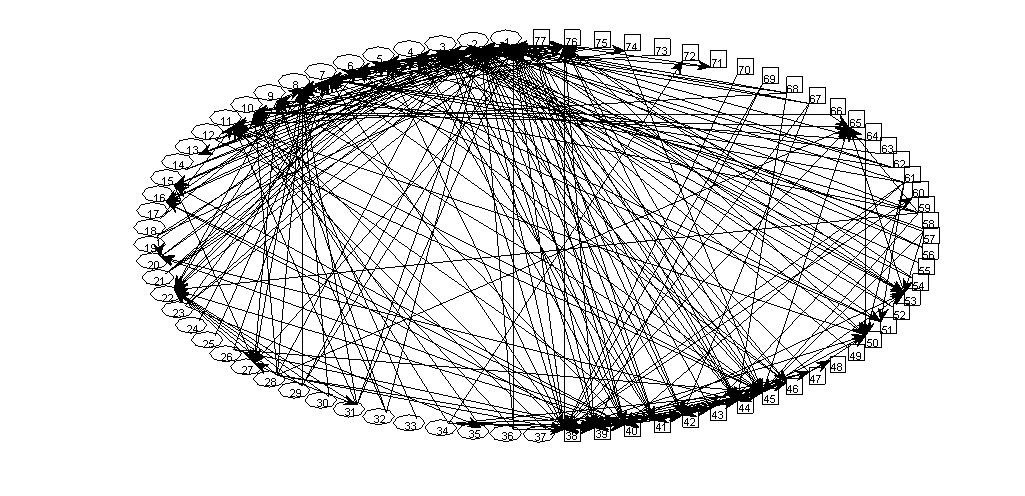}
\caption{ Network corresponding to the whole community  described by the $M_0$ matrix.  An arrow indicates a  directed link (DL). In contrast to Fig.1 no node is isolated } \label{fig:M0}
\end{figure}

  Figs. \ref{fig:C0D0}-\ref{fig:M0} display the corresponding $C_0$ and $D_0$, $A$,  $B$ and $M_0$, networks respectively.  Arrows indicate DL.  They point from $j$ to $i$, on an  $(i,j)$  link. There are
307 links, i.e. 102   among IDP, 86   among DED.  Among these we notice that there are  26 self-citations, leading to so called self-loops,  i.e. 11  in IDP, 15  in DED, i.e. amounting to a superfluous 11\% and 17\% contribution respectively, - they will be neglected here below.  It remains a total of 281 links, i.e. 91 in the IDP and 71 in the DED communities respectively, thus for a total of 162 (thus $\sim $ 57.7\%),   and 119 (thus $\sim $ 42.3\%)  inter-community links, indicating at once the relative importance of the inter-community exchanges.   Among these 89 are DL  (thus $\sim $ 74.8\%) and 15 are UL (thus $\sim $ 12.6\%) in other words    members from the other opinion community do not tend to go somewhat unnoticed, but most  arguing exchanges are rather concentrated among a few ($\simeq $ 15) opinion members. For completeness, let it be observed that
there are   219 DL, but only 31
UL  for the whole data set:  6  UL among IDP, 10 UL among DED, - an indication of internal community support, and a rough estimate of the number of opinion leaders; the other 15 UL concern cross-citing among the two groups as mentioned here above.  
  For further   reference we  report the detailed data in Table 1.

\begin{table}\label{tIDPDED}\begin{center} \begin{tabular}{|c|llll|}
\hline matrix & number   & trace &  &   \\
              &  of links &       &   DL & UL    \\
\hline $M $ & 307 &   26 &  219 &   31 \\
\hline $C$  &  102 &   11 &   79 &    6 \\
\hline $D  $&   86 &   15 &   51 &   10 \\
\hline $C_0 $&  91 &    0 &   79 &    6 \\
\hline $D_0 $&  71 &    0 &   51 &   10 \\
\hline $A  $&   86 &    0 &   86 &    0 \\
\hline $B$  &   33 &    0 &   33 &    0 \\
\hline $F$  &   119 &    0 &   89 &    15 \\
\hline $M_0$ &  281 &    0 &  219 &   31 \\

\hline \end{tabular}  \end{center}
\caption{Global description of the relevant matrices: trace, i.e. number of self-citations,
number of directed (DL) or undirected (UL) links. The total number of links is always for each matrix  equal to the number of directed links, plus the double of the number of undirected links, plus the trace (self-citations), i.e. the number of finite elements in the matrix}
\end{table}

\section{Data  statistical analysis } \label{sec:stat}

After   building the IDP and DED networks  and the
overall network of agents, due to links through the empirical
   observation outlined here above, we proceed performing some classical structural analysis on such
networks, i.e. an analysis of the node degree $k_i$ distribution, but taking into account   the directionality of the links  through the out-degree $k^{out}_i$ and in-degree  $k_i^{in}$ distributions, plus   indicative
coefficients, i.e. the network   assortativity, the (global and local) clustering coefficients and  the Average Overlap Index. 

\begin{figure}
\includegraphics[height=20cm,width=14cm]{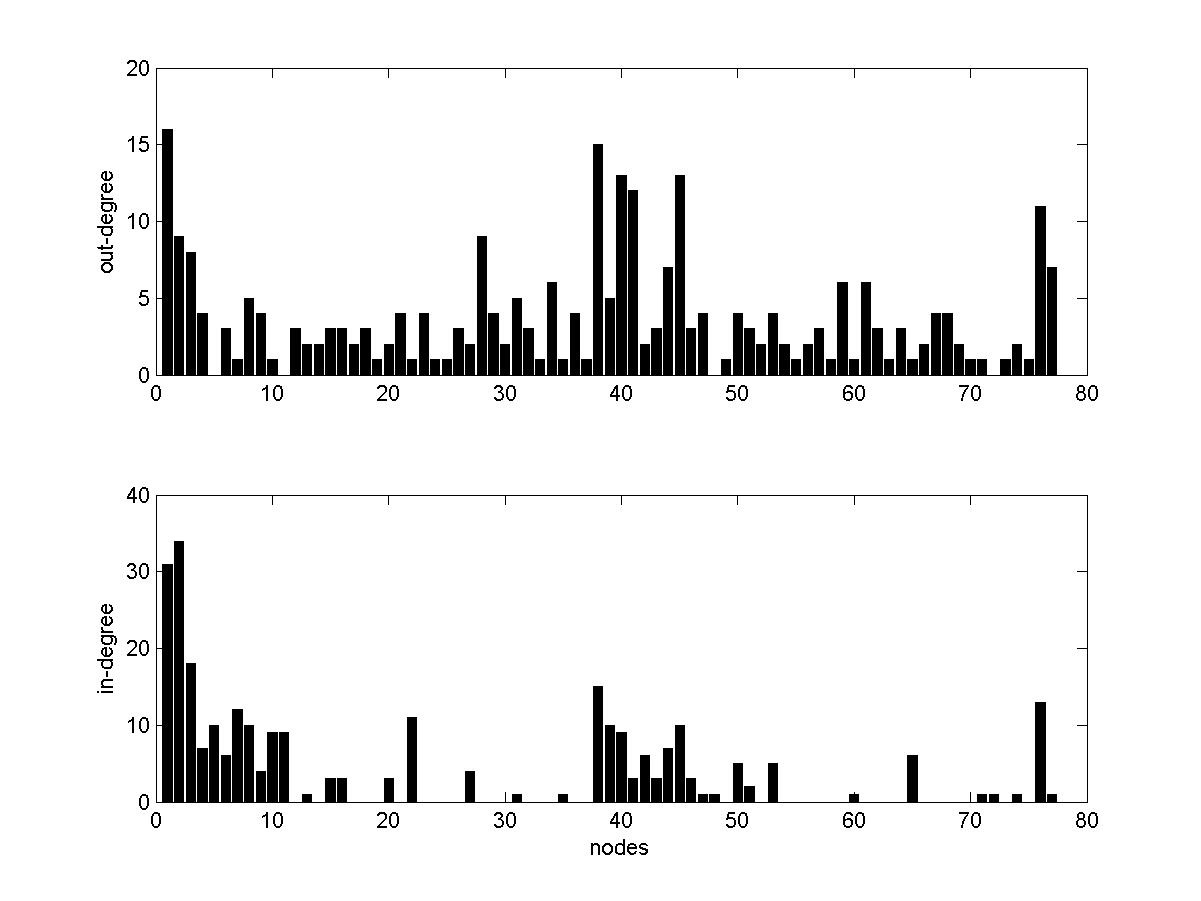}
\caption{Out$-$degree (top) and in-degree (bottom) for each node, from Matrix $M_0$.}
\label{fig:INOUTbar}
\end{figure}

\begin{figure}
\includegraphics[height=20cm,width=14cm]{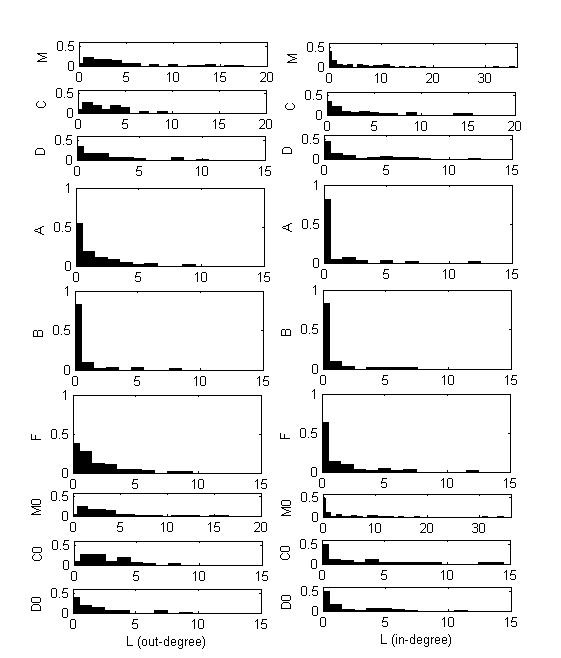}
\caption{Histogram of the number L of out$-$degrees  and in$-$degrees for each matrix recalled in  Table 1 }
\label{INOUTbar2height}
\end{figure}

\subsection{Degree distributions }

In this sub-section we report the results of the analysis of the empirical distributions of in-degree and out-degree, in order to be testing the hypothesis of power law (expected for "scale free networks") against exponential (expected for "random networks") distribution estimated on both probability density and distribution of links.   
Notice that a power law or exponential law, if found,  might also be meaningful for detecting the kind of growth of the network, respectively through a preferential attachment mechanism or through a random one.  One may also neither  know {\it a priori} whether different behaviors  exist or not for the out- and in-degree distribution, - moreover  depending on the matrix of interest.
Fig. \ref{fig:INOUTbar} shows, for each node, the out-degree and the in-degree, respectively, i.e. the number of links  exiting from  and entering into each node. Fig. \ref{INOUTbar2height} reports the  out- and in-degree histograms corresponding to the various matrices.
Finally notice that in doing so we add a quantitative set of values for an answer to a question raised in  \cite{amaralPNAS} on the classes of SWN so far examined in the literature.   

A power law behavior  is searched through a log-log plot method, i.e. the best fit of a power law  to the empirical density $p(x)$  of in-degree and out-degree for each matrix
\begin{equation}
\label{power}
       p(x) \sim x^{-b}
\end{equation}
where $x$ refers to the degree of a node, i.e. the number of links of the node.
The exponential law behavior is searched through the best fit of
\begin{equation}
\label{exp}
       p(x) \sim e^{-bx}
\end{equation}
to the empirical density $p(x)$  of  out-degree and in-degree for each matrix.

A gaussianity test of the residuals  has been performed for  cross-checking  the goodness of the fits
through  the Jarque-Bera (JB) test \cite{jbtest}.
The test returns the logical value 1 if
it rejects the null hypothesis at the 5 \% significance level, and 0 if
it does not. It can be concluded that the each fits is  statistically meaningful
when  the  Jarque-Bera (JB) test confirms the Gaussianity of the
residuals of $each$  fit in each case within the $95\%$ confidence bounds.
Table 2
reports the values of the parameter $b$ best fits  for the power law and  exponential law respectively for the empirical densities of out-degrees and in-degrees for each matrix/network of interest.

The behavior of the out-degree density is described $quasi$ equally well  by a power law  or by an exponential. The scale-free exponent of the out-degree of $M$ and its subnetwok is $b\sim1.0$  quite far from the scale free exponent of other communities, like the network of actors ($b=2.3$), or citation networks ($b \sim 3$) (see data  compendium in \cite{newman0303516}).
So mechanisms for out-links preferential attachment of  IDP-DED network, if any, are less significant than the one shown in other networks.
  The more so for the $b$ exponent   close to $b=0.5$ for $C$ and $C_0$, for the out-degree density.
  One might thus recognize that     the preferential attachment is much less likely than  for $M$ and  also $D$, or $M_0$ and also $D_0$, that show a higher value of $b$. However a value $b\sim4/3$ is found for the off-diagonal matrices $A$ and $B$. This can be interpreted as due to the fact that authors of different groups most refer to the most famous one of the other group, adding the strength of the outstanding author if compared to the others of his group.

   The comparison of the power law exponent of the in-degree cannot be performed because the JB test   rejects the Gaussianity of the residuals. Disregarding the JB test and accepting the exponent $b$ values there are found either close to 1.0 or 4/3. This would indicate a preference to citing a small group of authors for the in-degree density.

    The exponential law describes the in-degree decay much better than the power law. In fact, it is seen that the JB   test implies to reject the hypothesis of  the power law  behavior for the in-degree (Table 2). The power law is accepted only for $M$, for which also the  exponential decay is accepted. We can only conclude that in-degree of $M$ has a fast decay that is ''between'' the
power law and  the exponential. The prevailence of the exponential decay only signals that the small-world hypothesis may hold, but the rejection of the power law surely means that the network is not scale free, and that mechanisms of network growth  as through preferential attachment are unlikely to   occur.

  There are similarities in the range and hierarchy of values for the decay rate parameter $b$ of the out- and in-degree densities  for a given  matrix.   However a large difference is found in the relative values:  a factor of two frequently occurs.  The most marked difference is for the matrix $C$ and $C_0$,  having  $b\sim0.15$  and $b\ge0.75$  for the out- and in-degree respectively. Observe that the fact that the DED are more prone to cite themselves than the IDP much influences this $b$ value: it is very high $b\sim1.0$ for the former, very low for the latter $b\sim0.5$. This indicates either a marked narcistic or ego effect of the former community members, or an attitude in citing only respected or respectable scientists as in usual scientific publications \cite{citationnetworkredner,citationnetworkRadicchi}.
  This  fast decay has an interpretation however: it arises  from the
fact that there are very few persons citing or being cited by
many others, see the piling at low value of the degree on Fig. 6.

As a conclusion, let it be stated that we hardly observe any clear pattern of acceptance/rejection of  either empirical laws, nor thus on the type of network kinetics.
In fact, laws with tails should be considered to be very rough approximations of distribution functions when the finite size of the system is so much marked.

\begin{table} \hspace{-1cm}\label{tab:powerlaw}
\begin{tabular}{|c|llll|llll|} \hline
 &out-degree&  & out-degree  & & in-degree& & in-degree  & \\
\hline & $b$ (power law) & JB & $b$ (exp. law) & JB& $b$ (power law) & JB & $b$ (exp. law) & JB\\
\hline $ M$    & {\small 1.16 (0.88,1.44) } & {\small  1 } & {\small 0.13 (0.03,0.22) }  & {\small  0 } & {\small 0.72 (0.48,0.97) } & {\small  1 } & {\small 0.76 (0.54,0.98) } &       {\small  0} \\
\hline $ C$    & {\small 1.03 (0.21,1.85) } & {\small  0 } & {\small 0.12 (-0.11,0.34) }  & {\small  0 } & {\small 0.70 (0.38,1.01) } & {\small  0 } & {\small 0.50 (0.33,0.68) } &       {\small  0} \\
\hline $ D$    & {\small 0.75 (0.26,1.23) } & {\small  0 } & {\small 0.39 (0.21,0.57) }  & {\small  0 } & {\small 0.64 (0.15,1.13) } & {\small  0 } & {\small 0.83 (0.47,1.20) } &       {\small  0} \\
\hline $ A$    & {\small 1.31 (0.74,1.88) } & {\small  0 } & {\small 0.88 (0.65,1.11) }  & {\small  1 } & {\small 0.59 (0.06,1.12) } & {\small  0 } & {\small 2.83 (1.00,4.67) } &       {\small  0} \\
\hline $ B$    & {\small 0.68 (-0.62,1.99) } & {\small  0 } & {\small 2.20 (1.63,2.76) }  & {\small  0 } & {\small 1.00 (0.52,1.47) } & {\small  0 } & {\small 2.17 (1.81,2.54) } &       {\small  0} \\
\hline $ F$    & {\small 1.44 (1.14,1.74) } & {\small  0 } & {\small 0.48 (0.38,0.58) }  & {\small  0 } & {\small 0.96 (0.32,1.60) } & {\small  0 } & {\small 1.35 (0.96,1.75) } &       {\small  0} \\
\hline $M_0$    & {\small 1.26 (0.93,1.58) } & {\small  1 } & {\small 0.16 (0.03,0.29) }  & {\small  0 } & {\small 0.53 (0.20,0.86) } & {\small  0 } & {\small 1.26 (0.84,1.69) } &       {\small  0} \\
\hline $C_0$    & {\small 1.22 (0.39,2.06) } & {\small  0 } & {\small 0.15 (-0.15,0.44) }  & {\small  0 } & {\small 0.57 (0.19,0.95) } & {\small  1 } & {\small 1.17 (0.61,1.74) } &       {\small  1} \\
\hline $D_0$    & {\small 0.80 (0.27,1.33) } & {\small  0 } & {\small 0.48 (0.27,0.69) }  & {\small  0 } & {\small 0.63 (-0.03,1.29) } & {\small  0 } & {\small 0.99 (0.54,1.44) } &       {\small  0} \\
\hline
\end{tabular}
\caption{Test of  power law and of exponential decay for the empirical densities of out-degree and in-degree distributions for each discussed matrix.  The columns labelled JB report the results of the Jarque-Bera test, i.e. the test returns the logical value 1 if
 null hypothesis is rejected, and 0 if the null hypothesis is accepted. The  standard deviation confidence interval  is given in parentheses in each $b$ case}
\end{table}

\subsection{Network Assortativity}

In order to indicate some aspect of the attachment process in a network, one can calculate its so called assortivity. The term assortativity is commonly
used after \cite{assortativity,mixing} to refer to a preference for a network node to be attached to others depending on one out of many node properties. Assortativity is most often measured after a (Pearson) node degree correlation coefficient $r \in [-1,1]$, Eq.(\ref{r}) below, such that $r=1$ indicates perfect assortativity, $r=-1$ indicates perfect disassortativity, i.e. perfect negative correlation.
Let $p^{out}_i$  ($p^{in}_i$) be the probability that a randomly chosen vertex $i$ will have out-degree (in-degree)  $ k^{out}_i$ ($k^{in}_i$): they can be obtained/read from Fig. \ref{INOUTbar2height}.
Let $N$ be the number of nodes of the network, $L$ the number of links, and $ m_{ij}$ the adjacency matrix elements. By definition, adapting to our notations
 Eq.(26) of \cite{mixing}, the (network) assortativity coefficient $r$ is given by: 

\begin{equation}\label{r}  
r=\frac{\sum_{i,j=1}^N q^{in}_i q^{out}_j m_{ij}- (\sum_{i,j=1}^N q^{in}_i m_{ij})( \sum_{i,j=1}^N q^{out}_j m_{ij}) /L}{\sqrt{[\sum_{i,j=1}^N( q^{in}_i m_{ij})^2-(\sum_{i,j=1}^N q^{in}_im_{ij})^2/L][\sum_{i,j=1}^N (q^{out}_jm_{ij})^2 -(\sum_{i,j=1}^N q^{out}_jm_{ij})^2/L]}} 
\end{equation}

where
$$ q^{out}_i=\frac{k^{out}_i p^{out}_{i}}{\sum_i k^{out}_i  p^{out}_i};$$
and
$$  q^{in}_i=\frac{k^{in}_i p^{in}_{i}}{\sum_i k^{in}_i p^{in}_i}$$.

\begin{table}\label{tab:assortativity}
\begin{center}
\begin{tabular}{|l|l|l|l|l|l|l|l|l|l|l|} \hline
  &  $A$  & $B$  & $D_0$& $M$  & $M_0$ &  $F$& $D$ & $C_0$ & $C$ \\ \hline
$r$ & 0.253 & 0.397& 0.451 & 0.460 & 0.461 & 0.486 & 0.534& 0.621 & 0.644  \\ \hline
\end{tabular}
\caption{  Value of the assortativity coefficient Eq.(\ref{r}) following Eq.(26) of Newman \cite{mixing}.
The relevant matrices are ranked  hierarchily, from left (low $r$ value) to right (high $r$ value) indicating some increasing preferential attachment}
\end{center}
\end{table}

For the  networks of interest here, we have found,  see Table 3, quite positive values for the
$assortativity$,  ranging between 0.25 and 0.65.
    Again one can observe that the self-citations influence the $r$ value by 17\% in the case of DED. The $0\le r\le1$ values reflect the fact that there is
a small number of (to be later further precised as being more famous) authors either largely citing each other, or  largely cited  by each other, while the others
 show a low activity of citing. This is the case  for the DED, - matrix $D$.  The $r$ value for the whole network (matrix $M$ or $M_0$ is approximately the same as for the IDP, - matrix $C$ or $C_0$. Notice that $A$ and $B$ alone have the least  assortative   feature  than the other matrices (networks), as  could be expected from a psychological view point on this matter: the citations between rival groups tend  to be biased, i.e. they aim at    enhancing the most relevant authors/leaders of the rival community only \cite{newman,citationnetworkredner,citationnetworkRadicchi}.

\subsection{Clustering }\label{clustering}

The tendency of the network nodes to form local
interconnected groups is usually quantified by a measure
referred to as the clustering coefficient \cite{SWN}.
The amount of studies on this characteristic of networks has led to the particularization of the term to focus on different features of the networks. Here we consider the global clustering coefficient and the local clustering coefficient.

  Indeed, the most relevant elements of a heterogeneous agent interaction network can be identified by analyzing
  the global and  local connectivity properties. 
    In the present case those individuals leading the opinion of   IDP and DED groups
  can be attempted to be identified by analyzing the number of triangles with nodes belonging to the same community, so called homogeneous triangles,  or not,  so called inhomogeneous triangles,  beside the type of (directed or/and undirected) links of their citation network.  The former number gives some hierarchy information, the latter some reciprocity, i.e. recognition of leadership or proves of some challenging, conflict, in time.

 \subsubsection{Global clustering coefficient}
The global clustering coefficient (GCC) {\it of the network}
is defined as the average of $c_i$ {\it over all the vertices} in
the network, $<c>=\sum c_i /N$, where $N$ is the number of nodes of the network, and the
clustering coefficient $c_i$ of a vertex $i$ is given by the ratio
between the number $e_i$ of triangles  sharing that vertex,
and the  maximum number of triangles that the vertex could have. If a node $i$ has $k_i$ neighbors, then a clique, i.e.  a complete graph,  would have $k_i (k_i-1)/2 $ triangles at most, thus
$c_i=2e_i /k_i(k_i-1)$.  

     Whence one  starts by
     considering the various  triangles on the networks and next proceeds in calculating the
     clustering coefficient.
      This fact  should lead  to obtain two pools containing  respectively  the most important opinion leaders in the IDP and DED groups. However one
 has to generalize the    16  
  triplets nomenclature
 found in the literature,  \cite{pajek}   in order to consider the bi-community nature of the IDP-DED network, i.e. the node tag itself, beside the types of links, i.e. UL or DL, in between when they exist. 

\begin{figure} \begin{center}
\includegraphics[height=10cm,width=10cm]{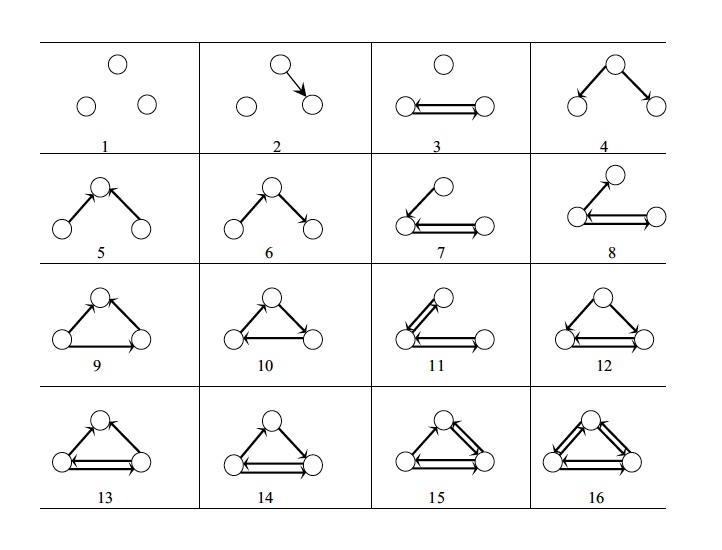}
\end{center}
 \label{3pajeklike} \caption{Pajek convention  \cite{pajek} Fig. 15, p. 52 for labelling 16 triplets of nodes variously linked.}
\end{figure}

       This leads to  consider  a  set of   104  different possible types of triplets (triangles if three nodes are linked) shown in Figs. 10-11 in Fig. 8 with appropriate notations  emphasizing the number of $C$ and $D$ nodes beside the number of DL and UL\footnote{Due to \cite{pajek} which listed triplets according to the number of UL and DL,  the case $\#11$ occurs when two nodes are  connected by two UL, thus have 4 links, -  a mere triangle which  has (at least) 3 links thus gets a lower $\#$ in \cite{pajek}. Thus triangles  are found in  patterns (9, 10) and (12, 13, 14, 15,  16). We have kept the \cite{pajek} order in our generalization. We apologize to the reader for this  apparently regrettable incoherence stemming from a different point of view between \cite{pajek}  and our aim. A similar comment  on misfit nomenclature can be made if one wishes to use the rank list of  (partially or not) connected vertices   in triplet cluster in \cite{milo}}. The counting  result is reported in Table 4. 

\begin{figure}\label{fig:2tri116}
\includegraphics[height=9cm,width=16cm]{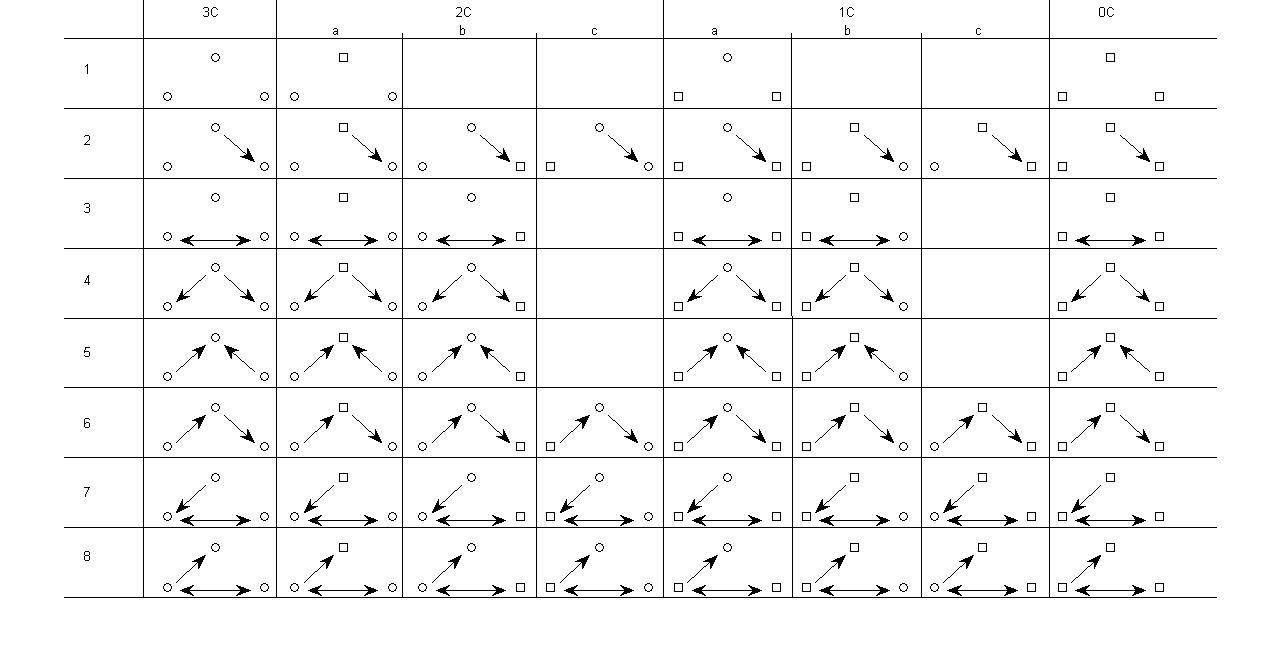}
\includegraphics[height=9cm,width=16cm]{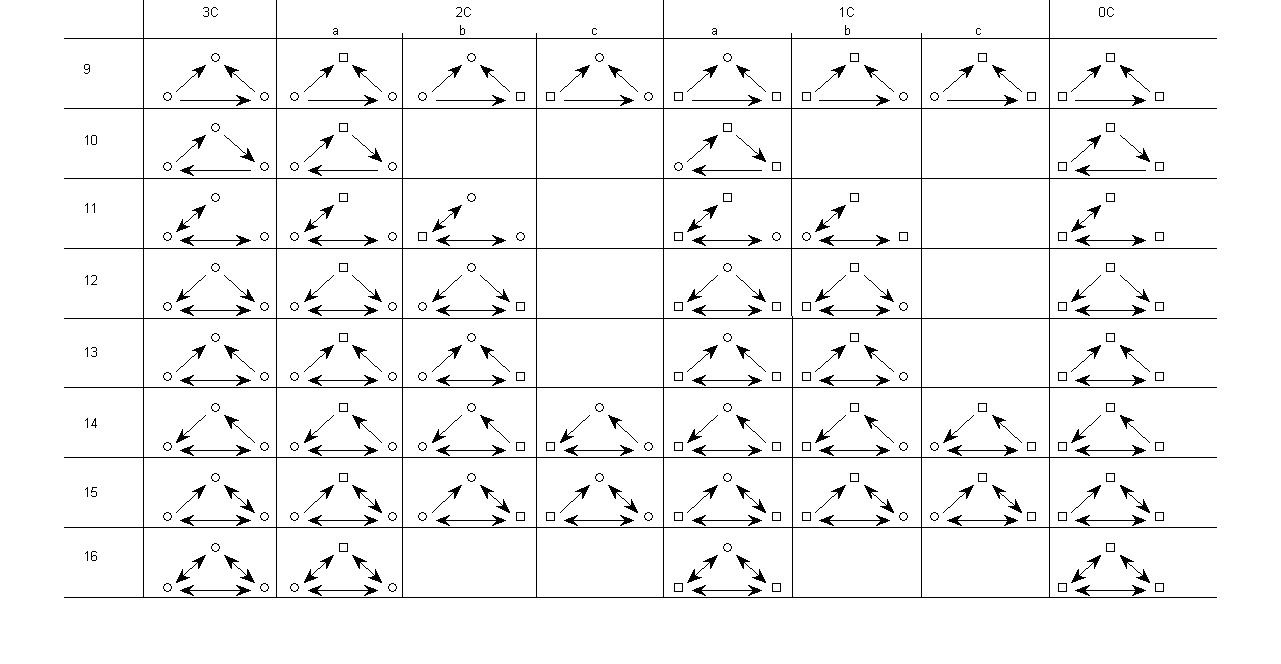}
 \caption{Present convention for labelling 104   triplets (triangles if three nodes are linked)  of nodes variously linked, taking into account the node nature:  a circle for  $creationists$ ($C$); a square for  $evolutionists$. 
The nomenclature of triangles with tagged vertices/nodes and biased edges/links has been generalized from the 16 combinations shown from the Pajek software  \cite{pajek}, that only consider the configurations shown in the first column. Such a generalization is necessary when   nodes, tagged  like  1/2-Ising  spins, belong to one out of two communities.
 Table 4 reports the triplets/triangles counted on the present case of IDP-DED opinions.
}
\end{figure}

\begin{table}\label{tab:total}
 \begin{center}
 \hspace{-2cm}
 \begin{tabular}{||c|c|c|c|ccc|c|ccc|c|c||}   \hline
triplet &  &3C  &  2C &  &   &  &  1C &   &   &   & 0C \\
type &  &  &    & 2Ca & 2Cb & 2Cc &   & 1Ca & 1Cb & 1Cc &   \\ \hline
1 &  57376 &   5320 &  20631 &  20631 &     -  &   -    &   23533 &  23533 &      - &      - &    7892 \\ \hline
2 &  11759 &   1874 &   4553 &    493 &   1708 &   2352 &    3888 &   1853 &    524 &   1511 &   1444 \\ \hline
3 &   1315 &    103 &    397 &    118 &    279 &      - &     563 &    267 &    296 &      - &    252 \\ \hline
4 &    240 &     40 &     75 &     47 &     28 &      - &      91 &      8 &     83 &     -  &     34 \\ \hline
5 &    963 &    185 &    453 &     15 &    438 &      - &     267 &    218 &     49 &      - &     58 \\ \hline
6 &    511 &    118 &    171 &     30 &     22 &    119 &     149 &     19 &     91 &    39  &     73 \\ \hline
7 &    489 &     52 &    206 &     65 &    121 &     20 &     186 &     33 &     20 &    133 &     45 \\ \hline
8 &    171 &     19 &     54 &      4 &     33 &     17 &      57 &     37 &     15 &      5 &     41 \\ \hline
9 &    104 &     33 &     26 &      1 &      6 &     19 &      33 &     29 &      0 &      4 &     12 \\ \hline
10 &     2 &      1 &      0 &      0 &      - &      - &       0 &      0 &      - &     -  &      1 \\ \hline
11 &    50 &      7 &     24 &     24 &      0 &      - &      16 &      2 &     14 &      - &      3 \\ \hline
12 &    13 &      0 &      7 &      0 &      7 &      - &       6 &      2 &      4 &      - &      0 \\ \hline
13 &    16 &      0 &      5 &      0 &      5 &     -  &       6 &      6 &      0 &      - &      5 \\ \hline
14 &    34 &      2 &      9 &      1 &      4 &      4 &      18 &      3 &      0 &     15 &      5 \\ \hline
15 &    42 &      2 &     15 &      1 &      1 &     13 &      18 &      9 &      9 &      0 &      7 \\ \hline
16 &    19 &      1 &      5 &      5 &      - &      - &      11 &     11 &      - &      - &      2 \\ \hline  \end{tabular}

\caption{  Column 1  lists  the type of  (connected or not) triplets as conventionally labelled in  the pajek software \cite{pajek} and recalled in Fig. 7 
.  Column 2 gives their number for the whole network of IDP-DED studied here.
The numbers in the other columns give the number of respective cases  according to the notations in Fig. 8 
.  Recall that triangles may include several different
$C$ and $D$  agents:  columns 3C, 2C, 1C, 0C, respectively refer to a triangle with 3
creationists, 2 creationists (and one darwinian evolutionist),
1 creationist (and 2 darwinian evolutionists), and 0 creationist (and 3 darwinian evolutionists). The letters (a,b,c)  serve to distinguish among the different configurations  shown in  Fig. 8; the order is arbitrary. }
 \end{center}
  \end{table}

 Notice at once that triangles containing elements/nodes  from $two$ groups are found to be the most abundant:  for the simplest triangle; $ \#$2C (=26) +$ \#$1C (=33)  $\ge$ $ \#$3C (=33) + $\#$0C (=12). Over all: there are 159 hybrid triangles, i.e. containing members from rival communities, for only 71 "homogeneous" triangles.  On the other hand several cases are completely missing: see many 0 values in   Table 4. Notice also that the number of triangles indicating DL $transitivity$\footnote{Let us use the symbol $\rightarrow$ for representing citations, i.e. $j \rightarrow  i $ means that  $j$ cites $i$.
 Let us assume that  $j \rightarrow  i $ and $i \rightarrow  k $. Then the relationship represented by $ \rightarrow  $ indicates a so called transitivity one if $j \rightarrow  k $   }, like $\#9$  are the most numerous ones, far more numerous than   those indicating a "round-and-around" $attack -citation$ pattern, i.e. $\#10$. There are even non-existent when heterogeneous triangles are considered.
 This {\it a priori}  unexpected fact immediately reflects a strongly peculiar interaction between both opinion groups.
 However when UL occur, the transitivity and attack-citation patterns are difficult to disentangle.  Cases  $ \#$15 and $ \#$14 are surely the most abundant ones in this case.

Unfortunately, one cannot easily explain from values on   both $ \#$15 and $ \#$14  (triplet type) data lines whether there is  some agent $support$ by its own community or some arguing by one or two of the same community against the other  along a similar line of arguments. In fact the measurements outlined till now cannot shed light on the relevance of a particular individual to  any network structure, because they provide a collective measure, whence do not focus on each single author.  This emphasises the need or  interest of looking at the $A$ and $B$ matrices more than at those on the diagonal, i.e. $C$ and $D$, as already proclaimed here above.

The detection of such triangles is at the base of our estimate of the global clustering coefficient $\Gamma$,
     defined as  the ratio between the number of triangles and the total number of possible triplets, 
 both numbers  which can  be  
easily obtained by summing the corresponding values listed in Table 4.   

Notice that $\Gamma$ can only be  unambiguously calculated for the as defined in \cite{pajek} configurations; otherwise a redistribution of the configurations has to be made to take into account symmetry and chirality  conditions among the possible104 configurations.

\begin{table}\label{table:Gammapajekonly}
\begin{center} 
 \begin{tabular}{llcllcl|c|lcl|c|lcll}  \hline
    \cite{pajek} notations&total     &3C        &  2C       & 1C 	& 0C \\ \hline
     $ \Gamma$  &   0.0867 &   0.0848 &   0.0638  &   0.1072 &     0.1119 \\ \hline
\end{tabular}
\end{center}
 \caption{Clustering coefficient, 
 i.e. the ratio of the number of  configurations of triangles, i.e. from  $ \#$9 to $ \#$16  (excluding the $ \#$11th) to the number of configurations implying at least three different nodes and two links, i.e.
$ \#$4  to $ \#$16  in  Fig. 7.}
 \end{table}

 \subsubsection{Local clustering coefficient}\label{subsectlcc}
In the literature  \cite{newman},   the term 'clustering coefficient' refers also to other quantities, relevant to understand the way in which nodes form communities, under some criterion.
By definition, the local clustering coefficient (LCC) $\Gamma_i$ for a node $i$ is the number of
 links between the vertices within the nearest neighbourhood of $i$ divided by the number of links that could possibly exist between them.
  It is relevant to note that the  GCC is not trivially related to the LCC, e.g. the GCC is not the mean of LCC.  In the former case,   triangles having common edges are emphasized, in the latter case only the links; the number of links common to triangles can vary much with the number of connected nearest neighbour nodes indeed.  Basically,  the latter quantifies how its neighbors are close to being part of a   complete graph.
LCC rather serves to determine whether a network is a SWN \cite{SWN} or not.

For a directed graph, for each neighbourhood of node $i$, there are at most $k_i(k_i - 1)$
links that could exist among the vertices within the neighbourhood, - where $k_i$ is the total (in- plus out-) degree of the vertex, i.e. $k_i=\sum_{j \neq i} (m_{ij}+m_{ji})$.
 Therefore\footnote{In the summations in the denominator of Eq.(7), and similarly for Eq.(8) and Eq. (9), we have not indicated the restriction $j\neq i$ , $i\neq k$  and  $k \neq j$ because  we consider the $M_0$, $C_0$, and $D_0$ matrices only}

\begin{equation}\label{lccformula}
\Gamma_i=\frac{k_i(k_i - 1)}{\sum_{j=1}^{N}(m_{ij}+m_{ji})+\sum_{k=j+1}^{N}\Upsilon(m_{ij}+m_{ji})\Upsilon(m_{ki}+m_{ik})(m_{jk}+m_{kj})}
\end{equation}

 where $\Upsilon$ is the $sign$ function, i.e. $\Upsilon(x)=0$ if $x=0$; $\Upsilon(x)=1$ if $x>0$; $\Upsilon(x)=-1$ if $x<0$.
 The first term in the denominator counts the links among $i$ and its neighbors. The second term in the denominator counts the links among the neighbors of $i$.
The lower limit for the index $k$ in   $\sum_{k=j+1}^{N}$ is introduced for avoiding double counting.

Considering only the links $exiting$  $i$, one could define a similar quantity  as follows
\begin{equation}\label{lccformulaOUT}
 \Gamma_i^{out}=\frac{k_i(k_i - 1)}{\sum_{j=1}^{N}(m_{ji})+\sum_{k=j+1}^{N}\Upsilon(m_{ji}) \Upsilon(m_{ki}) (m_{jk}+m_{kj})}
\end{equation}
 
or  considering only the links $entering$   $i$ as follows
\begin{equation}\label{lccformulaIN}
 \Gamma_i^{in}=\frac{k_i(k_i - 1)}{\sum_{j=1}^{N}(m_{ij})+\sum_{k=j+1}^{N}\Upsilon(m_{ij})  \Upsilon(m_{ik})(m_{jk}+m_{kj})}
\end{equation}

However, the $ \Gamma_i^{out}$ and $ \Gamma_i^{in}$  plots are not shown for space saving.

\begin{table}\label{table:LCC}
\begin{center} 
 \begin{tabular}{|l|c|c|c|c|c|c|}  \hline
                        &  $M_0$   &  $ C_0$   & $ D_0$  &  $A$     &   $ B$    & $ F$ \\ \hline
$\bar{\Gamma}$          &  0.411   &  0.387   &   0.321   & 0.216   &  0.124   & 0.245 \\ \hline
$max_i \Gamma_i$        &  0.750   &  0.667   &   0.667   & 0.500   &  0.500   & 0.500 \\ \hline
$min_i \Gamma_i$        &  0.117   &  0.140   &      0    &     0   &      0   &     0 \\ \hline \hline
$\bar{\Gamma}^s$        & 0.281    & 0.301    &   0.301   & 0.211   &  0.206   & 0.238 \\ \hline
$max_{i,s} \Gamma_i^s$  &  1.0000  &  1.0000  &  1.0000   & 0.5000  &  0.5000  &  1.0000 \\ \hline
$min_{i,s} \Gamma_i^s$  &   0      &     0    &     0     &    0    &    0     &   0 \\ \hline
\end{tabular}
\end{center}
 \caption{Local clustering coefficient, rounded to their third decimal,  of the relevant matrices calculated through Eq.(\ref{lccformula}) and  $\bar{\Gamma}=\frac{1}{N}\sum_i \Gamma_i$, where the sum is taken on the nodes of the network, and $N$ is the number of  nodes. Values are compared with the maximum and the minimum on the set of nodes. The bottom half of the table reports $\bar{\Gamma}^s$, i. e. the mean of the value $\bar{\Gamma}$ calculated on 1000 matrices obtained from the previous ones after shuffling.  $max_{i,s} \Gamma_i^s$ and $min_{i,s} \Gamma_i^s$  are, respectively, the maximum and the minimum taken over each $\Gamma_i$ of each simulation. Zeros are due to the fact that the shuffle may lead to isolated nodes.}
 \end{table}

Fig. 9 
  shows   the LCC $\Gamma_i$, both reporting the list of nodes on the x-axis and the same list sorted out according to the decreasing order of $\Gamma_i$,  while Fig. 10  
 is the  histogram of $\Gamma_i$.
These plots show that   there exists   a large  set of nodes having a  $\Gamma_i$  with zero value, and others, a very large number of triangles, mainly having a value close to 1/3 or 1/5.  Recall that the lower the $\Gamma$ value,  the less "fully connected" appears the network; such is the case of the  intra-community matrices, $A$ and $B$ and somewhat surprisingly $D_0$.

The local clustering coefficient  $\bar{\Gamma}$ for the whole system is given as the average  of the local clustering coefficient $\Gamma_i$ over all the nodes  \cite{SWN}, i.e. $\bar{\Gamma}=\frac{1}{N}\sum_i \Gamma_i$, where $N$ is the number of nodes.
The mean values for  the relevant matrices, as well as  the range through the min and max values, are reported in Table 6. 

Recall also here that a  graph is considered {\it small-world}, if its average local clustering coefficient  is significantly {\bf higher} than a random graph constructed on the same vertex set. Therefore this average   local clustering coefficient can be usefully compared with the mean value obtained by randomizing the network and its subnetworks.
Table  6
shows the results over 1000 random permutation of network links. The matrix $M$ has been shuffled before extracting again from it each submatrix ($M_0$, $C_0$, $D_0$, $A$, $B$, $F$).
For each selected matrix, let $\bar{\Gamma}(n)$ the value  $\bar{\Gamma}$ calculated after  the  $n-th$ shuffle. Then $\bar{\Gamma}^s=\frac{1}{1000}\sum_n \bar\Gamma(n)$.
  The values reported in Table 6  correspond to an average over 1000 cases of  the  respective $\Gamma$.

 It is seen that the mean of $\bar{\Gamma}$ of $M_0$ (0.41)  much higher than any of the mean of its random counterparts (0.28),   thereby indicating that the network is very far from a random graph, while the values inside each group are more close. This means that the intra-group interaction is quite close to a random one, but the build-up of inter-group relationship is far from being random.
Thus we confirm  that the present networks  look like SWN rather than RN.

\begin{figure}
\centering
\includegraphics[height=20cm,width=16cm]{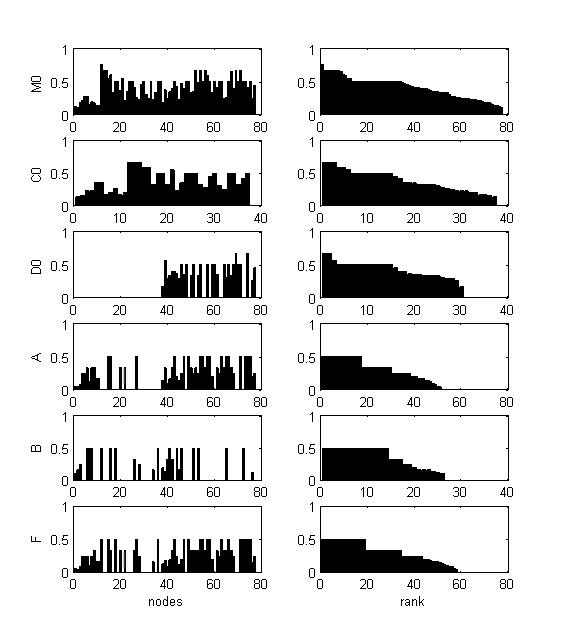}
\caption{Local clustering coefficients  ($LCC$)  $\Gamma_i$ as obtained from the $M_0$, $C_0$, $D_0$, $A$, $B$, $F$ matrices:
  (lhs): on the $x$-axis is the node number; (rhs):   the nodes ranked according to the decreasing value of  LCC on the $x$-axis   }
\label{fig:lcc2figs1}
\end{figure}

\begin{figure}
\centering
\includegraphics[height=20cm,width=16cm]{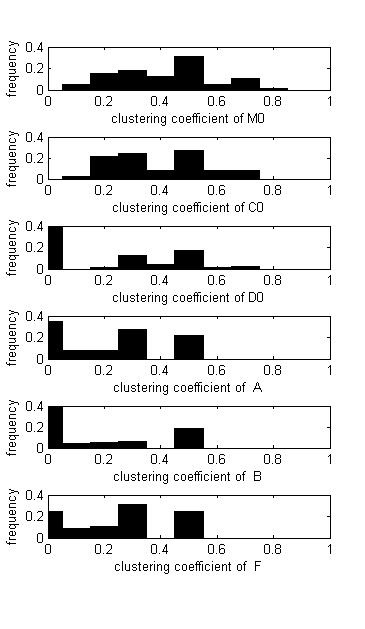}
\caption{Histograms of the local clustering coefficients ($LCC$)  $\Gamma_i$  as obtained from the  $M_0$, $C_0$, $D_0$, $A$, $B$, $F$  matrices respectively (top to bottom).
}
\label{fig:lcc2figs2}
\end{figure}

\begin{figure}
\centering
\includegraphics[height=18cm,width=15cm]{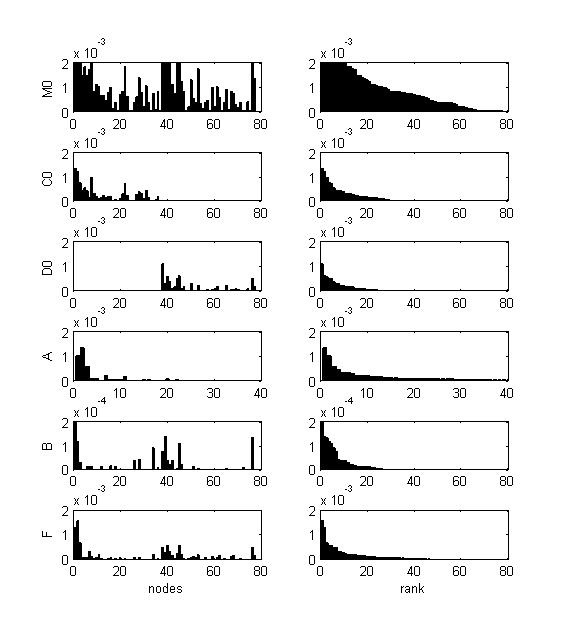}
\caption{ 
Average overlap index (AOI)    as obtained from the $M_0$, $C_0$, $D_0$, $A$, $B$, $F$ matrices: (lhs):  on the $x$-axis is the node number; (rhs):  on the $x$-axis, the nodes are ranked according to the decreasing AOI value  }
\label{fig:AOInew12}
\end{figure}

\subsection{Average Overlap Index}

Finally for characterizing members of communities, in another hierarchical way,  let us also calculate the Average Overlap Index (AOI) $O_{ij}$; its mathematical formulation and its properties are found in \cite{gligorausloos}
in the case of a unweighted network made of $N$  nodes linked by
 $(ij)$ edges,

\begin{equation}
\label{o} O_{ij} = \frac{{N_{ij} \left( {k_i + k_j }
\right)}}{{4\left( {N- 1} \right)\left({N - 2} \right)}},\quad
\quad i \ne j
\end{equation}
as before excluding self-citation  loops in calculating   $k_i$, and where $N_{ij}$ is the measure of the common number of neighbors to the
\textit{i}
and \textit{j} nodes. N.B., in a fully connected network, $N_{ij} = N - 2$. Of course, $O_{ii}=0$ by definition.

The \textit{Average Overlap Index} for the node   $i$ is
defined as
\begin{equation}  \label{average}
\left\langle {O_i} \right\rangle = \frac{1}{{N - 1}}\sum\limits_{j= 1}^N {%
O_{ij}. }
\end{equation}

 This measure, $\left\langle {O_i} \right\rangle$, can be interpreted indeed as an other  form of clustering attachment  measure: the higher the number of nearest neighbors, the higher the $\left\langle {O_i} \right\rangle$, the more so if the $i$ node has a high degree. Since the summation is made over  all possible $j$ sites connected to $i$ (over  all  sites in a fully connected graph), $\left\langle {O_i} \right\rangle$  expresses  a
measure of the local density near the $i$ site. In   magnetism,   the
links are like exchange integrals between spins located at $i$ and $j$.
That average over the exchange integrals is a measure of the critical
(Curie) temperature at which a spin system undergoes an order-disorder transition. Therefore $\left\langle             {O_i} \right\rangle$  can also  be interpreted as a measure of the stability of the node versus perturbations due to  a thermal  cause.  Thus, here,  a high $\left\langle {O_i} \right\rangle$  value reflects the $i$ node strong attachment to its community.

The \textit{average overlap index} of each
node, obtained according to Eq.(\ref{average}), depending on the matrix of interest
  are given in  Fig. 11. 
The order of magnitude of the  $\left\langle             {O_i} \right\rangle$ values are $\sim 10^{-3}$,  much smaller than in other investigated cases, like in  \cite{gligorausloos} or \cite{AOILA}. This is due to the low value of $N_{ij}$, somewhat reflecting the low GCC value for the whole network, i.e. 0.0867,  in Table 5., and the LCC values in Table 6.  Whence, a  rough estimation  suggests that  $\Gamma_i/N \; \sim O_i \sim 10^{-3}$, in good agreeemt indeed.

       \begin{table}\label{AOItops}
\begin{center} \begin{tabular}{|c|cl| }\hline
 agent  & node & AOI x $10^3$   \\
name&  number  & value\\ \hline
$M. Behe  $&2 &110 \\ \hline
$ W. Dembski $   &6&108 \\ \hline
$ S.C. Meyer$  &3 &54 \\ \hline
$C.R.  Thaxton $  &8&33 \\ \hline
.&. & .\\ \hline
$J. Murrel$ &33& 0\\ \hline
 \hline   \hline
$ R. Pennock$  &38 &69  \\ \hline
$ B. Forrest $  &40  &55 \\ \hline
$E.C. Scott $ &45  &52 \\ \hline
$ R. Dawkins $  &76 & 46 \\ \hline
$ E. Sober $  &44 & 35 \\ \hline
$ W. Elsberry $  &41 & 32 \\ \hline
$ K. Miller $ &39  &28\\ \hline
.&. & .\\ \hline
$M. Singham$ &72&0\\ \hline
\end{tabular}
\end{center} \caption{ IDP and DED agents ranked according to their AOI; we kept only those having an  AOI greater than 0.025  and the least  one for illustration }
\end{table}

   The AOI names and list  in Table 7 can be compared to the results on  the number of triangles of the most relevant actors.     It was noticed in \cite{garcia} that the IDP leaders had more homogeneous triangles (type $ \#$3C in the notations of Table 4   and Figs.7-8)   than others,  while the DED had (somewhat surprisingly) no triangle of type $ \#$0C, but IDP and DED agents had many "inhomegeneous" triangles.
They are
   \begin{itemize} \item
$M. Behe$, $ W. Dembski$,  $ S.C. Meyer$, and $C.R. Thaxton$;
  \item $ R. Pennock$,   $ B. Forrest$,   $E.C. Scott$,  $ R. Dawkins$, $ E. Sober$,
$ W. Elsberry$, and  $ K. Miller$.
\end{itemize} respectively.

\section{Conclusions}
\label{sec:conclusions}

 In primordial science, proto to scientific crises,
  there are
inter-connections between distinct disciplines which induce a
link,  between  authors  intra-connected otherwise through their
discipline, i.e. communities on networks. This is also displayed
in para-scientific disciplines, which leads to the temptation of
assimilating  either science to philosophy, metaphysics, religion or the opposite.  A debate attitude leads to a social behavior which has intuitively the same nature as  that in well founded scientific disciplines, i.e. a citation pattern to refute ideas or hypotheses, on one hand, and to claim some support from one's own community,
on the other hand.

In particular, the  creationism proposal and its subsequent sequel of propaganda
publications is reminiscent of the diffusion of   topics
in science. 
 This has  incited us into  examining this very modern case of  so called {\it scientific community} connectivities from a network of citations point of view. However community detection as in
\cite{ebeling,extra,PNAS101.04.5287,bernard}  is less relevant here from a statistical physics of network study  than the level of exchanges \cite{portercommunitystructurePhA386,communitystructurePorter0902,implicitcommIJMPC} and its reciprocity through arguing. The members of the communities are clearly belonging to one or another, as established by mere reading of their work, through arguments for and against others. Yet the inter-community, and at a lower level of interest the intra-community, links raise interesting network structural questions.

  In order to build the  network(s), we have applied the usual method \cite{newman},
namely we have considered a (citation) network of authors placed at nodes, with a link between them if they cite  another's paper.  E.g.  we have {\it a priori} discriminated two  sub-communities:  one for scientists favoring {\em "darwinian evolution"}  and others proposing {\em "creationism"} and {\it "intelligent design"} as a  scientific alternative. 
There were for the time of observation   (1990-2007)  of almost equal size, 37 and 40 agents respectively.

We have constructed the two networks and looked for their interaction through the number of undirected and directed links without weighting them nor discussing their chronological sequence.    In so doing it is found that the total number of directed links (219) is much higher than (31) that of undirected links, indicating an intense  debate on the controversy messages.
 There are  119 links between agents belonging to different communities,  respectively $86$ DL  in $A$,  i.e. DED citing IDP but only  $33$ DL in $B$, i.e. IDP citing DED.
Moreover this number of interconnections (119)  is quite superior to the corresponding link number  inside the  creationists  (91)  and the evolutionists (71) communities, indicating the  importance of the controversy exchanges.  Notice that the number of self-citations, in absolute values,  is greater  (15) for the true scientists than (11)   for the pseudo ones, - a  known  classical sin of that community.  The same observation goes  true in relative values.

The number of out-degrees  is much larger than the number of  in-degrees. Their distribution is however not fitting any obvious theoretical law, e.g. the JB test implies to reject the hypothesis
of the power law behavior for the in-degree. The  exponential decay-like law being more likely
indicates that  a small-world hypothesis can be imagined, but the rejection of the power
law means that the network is not scale free;  mechanisms of network
growth   through preferential attachment are unlikely to be the case.

There are similarities in the range and hierarchy of values for the  scale parameter   of the out- and in-degree densities for a given  network. A fast decay is found also indicating that the networks are not of the random type.  This has been confirmed when looking at the clustering coefficients resulting from   shuffled adjacency matrices and similarly grouping the nodes into two communities in Sec. \ref{subsectlcc}. 
  Amaral et al.  \cite{amaralPNAS} have proposed three classes of SWN: (i) scale-free networks, characterized by a vertex connectivity distribution that decays as a power law; (ii) broad-scale networks, characterized by a connectivity distribution that has a power law regime followed by a sharp cutoff; and (iii) single-scale networks, characterized by a connectivity distribution with a fast decaying tail. The   analyses presented in the main text and  summarized here above suggest that the IDP-DED networks belong to the the third  case.

In order to characterize the  necessarily small networks, based on a  adjacency matrices, we have  calculated  a few
{\em specialisation coefficients}.
One could consider other quantities of interest for networks \cite{bernard}, but some relevant point resides in the interestingly non symmetry of the citation networks.

The so called assortativity of the network has been examined in order to search whether there is a proof of  any preference of an agent  attachment to the sub-network of opposite (in contrast to same) opinion. Examining the whole network, the communities and the inter-community links it is found that  the agents are neither perfectly assortative nor perfectly  disassortative.
From the  the values of $r$, in Table   3, it is asserted that the networks are  weakly assortative. It is pointed out that
 the least
assortative features are in the inter-community networks,   as   expected if the citations between rival groups are biased, i.e. they aim at enhancing the most relevant
authors/leaders of the rival community.

In order to characterize in greater detail the intercommunity  structure and exchanges we have  considered elementary entities made of agents belonging to different communities. The smallest  (geometric) cluster to be examined is the triangle. In order to do so we have generalized the usual nomenclatures of  triangles in order to take into account a specific tag to the nodes beside the number of links between these tagged nodes.

We have distinguished between  closed triangles and  triplets, - in which only two  neighbors of  a node are  $not$ otherwise connected. Thus we have calculated  the global clustering  coefficients, searching for the most relevant triangles.
It can be noticed that the number of triangles
indicating transitivity, like  $\#9$, in the nomenclature (Figs. 7-8) is quite higher than the number of those indicating
a round-and-around pattern, i.e.  like $\#10$ are completely missing. This {\it a priori} unexpected fact immediately reflects a strongly peculiar, very direct interaction between both opinion groups, i.e.  a weakly collective dynamics.
The transitivity patterns being more numerous  in $\#$1C than in $\#$2C  cases  seem also to indicate more agent support by its own (DED) community or some arguing   into a similar line. This emphasises the interest of looking at the $A$ and $B$ matrices more than at those on the diagonal, as proclaimed since the Introduction section.

 An aspect of these sub-networks, or communities  should be strongly emphasized. Although the number of triangles involving members of the different  communities is very large and approximately equivalent, see above, it is found at this stage that there is apparently no homogeneous triangle involving the leaders of opinion if they belong to the DED community. We attribute this to a stronger rivalry in the DED community than in the IDP ideology.
This is {\bf hardly seen}  if the analyses had been  stopping at a tag-free set of agents,   thus   is an interesting argument in favor of examining networks with tagged nodes and biased links  in further works.

 In this respect the study of the  local clustering  coefficients  indicates  a low value for the inter-community networks.
 The above results  indicate some leading opinion  leaders or  rather controversy makers. The average overlap index (AOI) \cite{gligorausloos}  allows to extract  from the  clusters those persons which  inside their community and with respect to the other are the centers of more attention, i.e. see Table 8.

Comments and suggestions on modelization of such society structure can be now in order. One may first consider the practical aspects resulting from the node characteristics, next those from links. The main opinion models in the literature consider the node state as somewhat independent of the link state, merely carrying some information flow.  The basic question is whether some stable phase exists and whether some phase transition can take place and if so under what circumstances, geometric, or "energetic" condition. from a

 Since the networks are not scale free, nor random, since nevertheless  there is some attachment effect to the community
the spreading epidemics or its confinement should occur  through a change in opinion, not of the hub (inflexible) nodes or agents, but of the agents at the  border of one of the communities.  This does not contradict Galam \cite{galamdebates}  assertion that    to focus on convincing open
mind agents is useless when there are  so claimed incomplete or dubious data. Galam observes that {\it when the scientific evidence is not as
strong as claimed, the inflexibles} (in whatever community) {\it rather than the data are found to drive the
collective opinion of the population}. Whence a strong emphasis on node roles.

 Practically this suggests one way to soften the controversy. {\it To produce inflexibles in one's own side is thus critical to win a public argument whatever the rigor cost and the
associated epistemological paradoxes }\cite{galamdebates}.

Thus conducting a thorough analysis of (both) issues  and to adopt a fair discourse are   lose-out
strategy  ingredients.   Adopting  a cautious balanced attitude based on whatever scientific or not facts in contrast with an attitude of overstating arguments and asserting wrong statements, -  which cannot be scientifically or {\it a priori}  disregarded, is found to be necessary  to eventually win a public debate. This is of interest in a democratic like system or meeting.   However we conjecture that in a system in which  the "votes" are taken over a long time scale, when the hubs are the inflexibles, as here, while the other agents have  loose bounds, small connexions, and belong to heterogeneous triplets  the change of opinion might occur. It is not obvious that  {\it to adopt a cynical behavior is the obliged path to win a public debate against unfair and rigid opponents.} Indeed, how long would remain an opponent convinced of the rightness of the other attitude when he/she perceives the lies.

 In view of the analysis and values reported in Table 1, it appears that there are 104 intercommunity links out of 526, whence $\simeq$ 0.20 of the total amount, i.e. a proportion above the critical value 0.15 found in \cite{majoritymodel}.

However, qualitatively, one can expect that the dynamics is driven by the collaboration and the intercommunity  links, and not by  a degree of freedom, like a "spin", attached to a node. Thus a model taking into account the number of interconnexions  \cite{1JSM,majoritymodel} would seem to be more appropriate. Our analysis indicates that there are  20\%  intercommunity link, -s much above  the critical value 0.15 for which a transition to an ordered consensus state is possible according to the model in \cite{1JSM,majoritymodel}.

 It is intuitively obvious that there is no ''phase transition''  to be expected in the present system. Indeed the communities are pretty much behaving as in a very deep {\it potential well}.  Their opinion can hardly be modified even through increasing the number of links or the intensity of the information exchange. These might even be counterproductive \cite{galamdebates}. In other words, there is not much change/fluctuations in the node state, whatever the link weight or node degree. It can surely be  said that  a  phase state change is likely more   ''easy'' in economy and politics than in  so called religious matter, - assuming a logical or scientific approach.

Therefore a third type of model comes in mind: the importance of the links in defining the node state and community makes us think of an analogy with ferro-electric models in Statistical Mechanics.  Remember that the links carry a sense in such models, and the site weight depends on the sense of the arrows on the links out- or in-coming on the site. In so doing the interaction energy between two nodes depends on the configuration of these links at both connected nodes. It is thereby suggested to discuss where IDP-DED stability points defined through an order parameter connected to the set of triangles can be found  in a phase diagram;   the empirical observation of the emergence of homogeneous  phases in large bi-community  networks might help in at first a mean field approximation approach.

 In more sophisticated  models  one could take into account that an opinion is not
simply  +1 and -1, as in so  called "categorical" ones, but rather consider that the tag  can take discrete values, i.e., with several "levels" as also suggested in  \cite{CHESS}, i.e. being a so called "ordinal"  or continuous-like variable attached at a node \cite{categorical}.

\bigskip

{\bf Acknowledgements} MA  would like to thank  A. Garcia Cantu
Ross for previous discussions \cite{garcia}.  MA acknowledges  FNRS  FC 4458 - 1.5.276.07 F project having allowed some stay at U. Tuscia. We thank the COST Action MP0801 for financial support, in particular of GR, through the STSM 4475.

\newpage
\section*{Appendix A: Historical perspective}\label{sec:App_history}

Let us briefly cite \cite{garcia} to put some historical perspective: {\it One may distinguish two main
opinion groups  about the subject of the origin of the universe
and life. The first group holds the $scientific$ consensus and in
particular Darwin  evolution theory as a valid basis, whilst the
second  is formed by people adopting a $theistic$ (in the present
context, biblical) view where natural processes are conceived as
occurring out of the purposeful will of some supra natural entity.
Amid them some belong to a historical movement, so called
$creationism$, which aims to refute and overturn Darwin theory.
Such organized opposition to evolution has found most of its
adherents in different Christian traditions, actively engaged in
promoting their values at the core of society. The influence of
some of these religious groups has been especially relevant in
some regions of USA and is introduced in other parts of the world,
for reasons not discussed at this level.}  and {\it  Specifically,
the ID proponents  $(IDP)$ have employed concepts of information
theory, thermodynamics and molecular biology in seeking for
evidences of an intelligent blueprint underlying the complexity
observed in biological systems. Yet, none of the IDP arguments has
been validated by most of the scientific community. In spite of
this fact, the ID movement has further developed: since the second
half part of the 1990Õs, in good measure thanks to the support and
headquarter brought by the Discovery InstituteÕs Center for
Science and Culture (CSC)}\cite{CSC}.

The increasing activity and impact of the ID movement has impelled
the reaction of social and scientific organizations around the
world. Among the most important ones, the non-profit organization
National Center for Science Education (NCSE) \cite{NCSE} plays a
relevant role in coordinating the activity of people defending the
teaching of evolutionary biology in the USA. Hereafter we refer to
the international group of people fighting ID as Darwin evolution
defenders (DED).

\bigskip

\section*{Appendix B: IDP-DED matrix}\label{sec:App_matrix}

In this Appendix we give the adjacency matrices of interest: $C$, $A$, $B$, $D$.

\begin{table}\label{table:C} 
\begin{tabular}{|l|l|l|l|l|} \hline $C$ &{\small  1 \hspace{4pt} $\cdots$\hspace{5pt} 5 \hspace{4pt}$\cdots$ \hspace{5pt}9} &{\small 10 \hspace{3pt}  $\cdots$ \hspace{3pt}15 \hspace{3pt}$\cdots$\hspace{3pt} 19 }& {\small 20 \hspace{4pt} $\cdots$ 25 \hspace{4pt} $\cdots$ 29 }& {\small 30 \hspace{11pt} $\cdots$ \hspace{11pt} 37 } \\ \hline
 1& 1  1  1  1  \hspace{1.1pt}$\cdot$\hspace{1pt}  \hspace{1.1pt}$\cdot$\hspace{1pt}  \hspace{1.1pt}$\cdot$\hspace{1pt}  \hspace{1.1pt}$\cdot$\hspace{1pt}  \hspace{1.1pt}$\cdot$\hspace{1pt} & 1  \hspace{1.1pt}$\cdot$\hspace{1pt}  1  1  1  1  1  \hspace{1.1pt}$\cdot$\hspace{1pt}  \hspace{1.1pt}$\cdot$\hspace{1pt}  \hspace{1.1pt}$\cdot$\hspace{1pt} & \hspace{1.1pt}$\cdot$\hspace{1pt}  1  \hspace{1.1pt}$\cdot$\hspace{1pt}  1  \hspace{1.1pt}$\cdot$\hspace{1pt}  \hspace{1.1pt}$\cdot$\hspace{1pt}  \hspace{1.1pt}$\cdot$\hspace{1pt}  \hspace{1.1pt}$\cdot$\hspace{1pt}  \hspace{1.1pt}$\cdot$\hspace{1pt}  1 & \hspace{1.1pt}$\cdot$\hspace{1pt}  1  \hspace{1.1pt}$\cdot$\hspace{1pt}  \hspace{1.1pt}$\cdot$\hspace{1pt}  \hspace{1.1pt}$\cdot$\hspace{1pt}  \hspace{1.1pt}$\cdot$\hspace{1pt}  1  \hspace{1.1pt}$\cdot$\hspace{1pt} \\
\hline
\end{tabular} \caption{Matrix C}\end{table}

\newpage
\begin{table}\label{table:D} 
\begin{tabular}{|l|l|l|l|l|} \hline $D$&{\small  38 \hspace{4pt} $\cdots$\hspace{5pt} 43 \hspace{4pt}$\cdots$ \hspace{5pt}47} &{\small 48 \hspace{3pt}  $\cdots$ \hspace{3pt}53 \hspace{3pt}$\cdots$\hspace{3pt} 57 }& {\small 58 \hspace{4pt} $\cdots$ 63 \hspace{4pt} $\cdots$ 67 }&{\small 68 \hspace{4pt} $\cdots$ 73 \hspace{4pt} $\cdots$ 77 }\\ \hline
38& 1  1  1  1  \hspace{1.1pt}$\cdot$\hspace{1pt}
\hspace{1.1pt}$\cdot$\hspace{1pt}  1  1
\hspace{1.1pt}$\cdot$\hspace{1pt}  1 &
\hspace{1.1pt}$\cdot$\hspace{1pt}
\hspace{1.1pt}$\cdot$\hspace{1pt}
\hspace{1.1pt}$\cdot$\hspace{1pt}
\hspace{1.1pt}$\cdot$\hspace{1pt}
\hspace{1.1pt}$\cdot$\hspace{1pt}
\hspace{1.1pt}$\cdot$\hspace{1pt}
\hspace{1.1pt}$\cdot$\hspace{1pt}
\hspace{1.1pt}$\cdot$\hspace{1pt}
\hspace{1.1pt}$\cdot$\hspace{1pt}
\hspace{1.1pt}$\cdot$\hspace{1pt} &
\hspace{1.1pt}$\cdot$\hspace{1pt}
\hspace{1.1pt}$\cdot$\hspace{1pt}
\hspace{1.1pt}$\cdot$\hspace{1pt}  1
\hspace{1.1pt}$\cdot$\hspace{1pt}
\hspace{1.1pt}$\cdot$\hspace{1pt}
\hspace{1.1pt}$\cdot$\hspace{1pt}
\hspace{1.1pt}$\cdot$\hspace{1pt}
\hspace{1.1pt}$\cdot$\hspace{1pt}  1 &
\hspace{1.1pt}$\cdot$\hspace{1pt}  1  1
\hspace{1.1pt}$\cdot$\hspace{1pt}
\hspace{1.1pt}$\cdot$\hspace{1pt}
\hspace{1.1pt}$\cdot$\hspace{1pt}
\hspace{1.1pt}$\cdot$\hspace{1pt}
\hspace{1.1pt}$\cdot$\hspace{1pt}  1
\hspace{1.1pt}$\cdot$\hspace{1pt} \\
\hline
\end{tabular}\caption{Matrix D} \end{table}
\newpage
\newpage
\begin{table}\label{table:A} 
\begin{tabular}{|l|l|l|l|l|} \hline $A$&{\small  38 \hspace{4pt} $\cdots$\hspace{5pt} 43 \hspace{4pt}$\cdots$ \hspace{5pt}47} &{\small 48 \hspace{3pt}  $\cdots$ \hspace{3pt}53 \hspace{3pt}$\cdots$\hspace{3pt} 57 }& {\small 58 \hspace{4pt} $\cdots$ 63 \hspace{4pt} $\cdots$ 67 }&{\small 68 \hspace{4pt} $\cdots$ 73 \hspace{4pt} $\cdots$ 77 }\\ \hline
 1& 1  1  1  1  1  1  1  1  \hspace{1.1pt}$\cdot$\hspace{1pt}  \hspace{1.1pt}$\cdot$\hspace{1pt} & \hspace{1.1pt}$\cdot$\hspace{1pt}  \hspace{1.1pt}$\cdot$\hspace{1pt}  \hspace{1.1pt}$\cdot$\hspace{1pt}  \hspace{1.1pt}$\cdot$\hspace{1pt}  1  1  \hspace{1.1pt}$\cdot$\hspace{1pt}  \hspace{1.1pt}$\cdot$\hspace{1pt}  1  \hspace{1.1pt}$\cdot$\hspace{1pt} & \hspace{1.1pt}$\cdot$\hspace{1pt}  \hspace{1.1pt}$\cdot$\hspace{1pt}  \hspace{1.1pt}$\cdot$\hspace{1pt}  \hspace{1.1pt}$\cdot$\hspace{1pt}  1  \hspace{1.1pt}$\cdot$\hspace{1pt}  1  \hspace{1.1pt}$\cdot$\hspace{1pt}  \hspace{1.1pt}$\cdot$\hspace{1pt}  1 & 1  \hspace{1.1pt}$\cdot$\hspace{1pt}  \hspace{1.1pt}$\cdot$\hspace{1pt}  \hspace{1.1pt}$\cdot$\hspace{1pt}  \hspace{1.1pt}$\cdot$\hspace{1pt}  \hspace{1.1pt}$\cdot$\hspace{1pt}  \hspace{1.1pt}$\cdot$\hspace{1pt}  \hspace{1.1pt}$\cdot$\hspace{1pt}  1  1 \\ \hline
 2& 1  1  1  1  1  \hspace{1.1pt}$\cdot$\hspace{1pt}  1  1  1  1 & \hspace{1.1pt}$\cdot$\hspace{1pt}  1  1  1  \hspace{1.1pt}$\cdot$\hspace{1pt}  1  \hspace{1.1pt}$\cdot$\hspace{1pt}  \hspace{1.1pt}$\cdot$\hspace{1pt}  \hspace{1.1pt}$\cdot$\hspace{1pt}  \hspace{1.1pt}$\cdot$\hspace{1pt} & \hspace{1.1pt}$\cdot$\hspace{1pt}  1  \hspace{1.1pt}$\cdot$\hspace{1pt}  1  1  \hspace{1.1pt}$\cdot$\hspace{1pt}  \hspace{1.1pt}$\cdot$\hspace{1pt}  \hspace{1.1pt}$\cdot$\hspace{1pt}  \hspace{1.1pt}$\cdot$\hspace{1pt}  1 & 1  \hspace{1.1pt}$\cdot$\hspace{1pt}  \hspace{1.1pt}$\cdot$\hspace{1pt}  1  \hspace{1.1pt}$\cdot$\hspace{1pt}  \hspace{1.1pt}$\cdot$\hspace{1pt}  \hspace{1.1pt}$\cdot$\hspace{1pt}  \hspace{1.1pt}$\cdot$\hspace{1pt}  1  1 \\
\hline
\end{tabular} \caption{Matrix A} \end{table}

\begin{table}\label{table:B} \begin{tabular}{|l|l|l|l|l|} \hline $B$&{\small  1 \hspace{4pt} $\cdots$\hspace{5pt} 5 \hspace{4pt}$\cdots$ \hspace{5pt}9} &{\small 10 \hspace{3pt}  $\cdots$ \hspace{3pt}15 \hspace{3pt}$\cdots$\hspace{3pt} 19 }& {\small 20 \hspace{4pt} $\cdots$ 25 \hspace{4pt} $\cdots$ 29 }& {\small 30 \hspace{11pt} $\cdots$ \hspace{11pt} 37 } \\ \hline
38& 1  1  \hspace{1.1pt}$\cdot$\hspace{1pt}
\hspace{1.1pt}$\cdot$\hspace{1pt}
\hspace{1.1pt}$\cdot$\hspace{1pt}
\hspace{1.1pt}$\cdot$\hspace{1pt}
\hspace{1.1pt}$\cdot$\hspace{1pt}
\hspace{1.1pt}$\cdot$\hspace{1pt}
\hspace{1.1pt}$\cdot$\hspace{1pt} &
\hspace{1.1pt}$\cdot$\hspace{1pt}
\hspace{1.1pt}$\cdot$\hspace{1pt}
\hspace{1.1pt}$\cdot$\hspace{1pt}
\hspace{1.1pt}$\cdot$\hspace{1pt}
\hspace{1.1pt}$\cdot$\hspace{1pt}
\hspace{1.1pt}$\cdot$\hspace{1pt}
\hspace{1.1pt}$\cdot$\hspace{1pt}
\hspace{1.1pt}$\cdot$\hspace{1pt}
\hspace{1.1pt}$\cdot$\hspace{1pt}
\hspace{1.1pt}$\cdot$\hspace{1pt} &
\hspace{1.1pt}$\cdot$\hspace{1pt}
\hspace{1.1pt}$\cdot$\hspace{1pt}
\hspace{1.1pt}$\cdot$\hspace{1pt}
\hspace{1.1pt}$\cdot$\hspace{1pt}
\hspace{1.1pt}$\cdot$\hspace{1pt}
\hspace{1.1pt}$\cdot$\hspace{1pt}
\hspace{1.1pt}$\cdot$\hspace{1pt}
\hspace{1.1pt}$\cdot$\hspace{1pt}  1
\hspace{1.1pt}$\cdot$\hspace{1pt} &
\hspace{1.1pt}$\cdot$\hspace{1pt}
\hspace{1.1pt}$\cdot$\hspace{1pt}
\hspace{1.1pt}$\cdot$\hspace{1pt}
\hspace{1.1pt}$\cdot$\hspace{1pt}
\hspace{1.1pt}$\cdot$\hspace{1pt}
\hspace{1.1pt}$\cdot$\hspace{1pt}  1
\hspace{1.1pt}$\cdot$\hspace{1pt} \\
\hline
\end{tabular} \caption{Matrix B}\end{table}


\begin{thebibliography}{0}

\bibitem{stauffer1} D. Stauffer, 
  Physica A  {\bf 336} (2004) 1-5.

\bibitem{pastor}
R. Pastor-Satorras and A. Vespignani, {\em Evolution and Structure
of the Internet : A Statistical Physics Approach}, (Cambridge
University Press, Cambridge, 2004); R. Pastor-Satorras, M. Rubi,
and A. Diaz-Guilera, {\em   Statistical Mechanics of Complex
Networks}, Lect. Notes Phys. {\bf 625} (Springer, Berlin, 2003)

\bibitem{1JSM} R. Lambiotte and M. Ausloos,
 J. Stat. Mech.: Theor. Exp. (2007) P08026


 \bibitem{econophysics} {\it Empirical sciences of financial fluctuations. The advent of econophysics}, Tokyo, Japan, Nov. 15-17, 2000 Proceedings H. Takayasu, Ed. (Springer Verlag, Berlin, 2002) ; {\it Practical Fruits of Econophysics}, H. Takayasu, Ed., (Springer, Tokyo,  2006).
 

   \bibitem{sociophysics} {\it Econophysics and Sociophysics}, B.K.  Chakrabarti, A. Chakraborti and A. Chatterjee,  Eds. (Wiley-VCH, Weinheim, 2006).


\bibitem{costa} L.F. Costa, O.N. Oliveira, G. Travieso, A. Rodrigues, P.R. Villas Boas,
L.  Antiqueira, M.P. Viana, and L.E. Correa da Rocha,
 {\it Analyzing and Modeling Real-World Phenomena with Complex Networks: A Survey of Applications},
arXiv:0711.3199v2 

        \bibitem{APE98}  A.  P\c{e}kalski,
        Physica A   {\bf 252} (1998) 325-335.

     \bibitem{pnas.99.02.7821hub} M. Girvan and M.E.J. Newman,
     {\em Proc. Natl. Acad. Sci. USA} {\bf 99} (2002) 7821-7826.

      \bibitem{PNAS-Netw-04slanina} M. Marsili, F.  Vega-Redondo, and F. Slanina,
      {\em Proc. Natl. Acad. Sci. USA} {\bf 101} (2004) 1439-1442.

\bibitem{milgram}  S. Milgram, Psychol. Today  {\bf 1} (1967)  60-67.

\bibitem{killworth} P. D. Killworth, E.C. Johnsen, H. R. Bernard, G.A. Shelley, and C. McCarty,
Social Networks  {\bf 23} (1990) 289-312. 

\bibitem{majoritymodel} R. Lambiotte, M. Ausloos, and J.A. Holyst, 
Phys. Rev. E  {\bf 75}  (2007)  030101(R). 

\bibitem{lambiotte1} R. Lambiotte and P. Panzarasa,
J.  Informetrics   {\bf 3} (2009) 180-190.


\bibitem{lambiotte2}  T. Evans and R. Lambiotte,
Phys. Rev. E  {\bf 80} (2009) 016105.

\bibitem{froncszak} P. Fronczak, A. Fronczak, and J. A. Holyst,  Phase transitions in social networks,
$http://arxiv.org/abs/physics/0701182$

\bibitem{vicsek} I. Derenyi, I. Farkas, G. Palla, and T. Vicsek,
{\em Phys. Rev. E} {\bf 69}  (2004)    046117.

\bibitem{SWN} D.J. Watts and S.H. Strogatz, Nature  {\bf 393}  (1998)   440-442.

  \bibitem{svenson02} P. Svenson and D.A. Johnston,
  {\em Phys. Rev. E} {\bf 65} (2002) 036105. 

   \bibitem{religion1}  M. Ausloos and  F. Petroni,
   Europhys. Lett. {\bf 77}  (2007) 38002. 


\bibitem{dv00}
Z. I. Dimitrova and N. K. Vitanov,
{\it Phys. Lett. A} {\bf 272} (2000) 368-380.

\bibitem{steele2} C. Buckley and J. Steele,
{\it World Archaeology} 
{\bf 34}  (2002) 26-46.


    \bibitem{PNAS99.02.7267_moss} S. Moss,
    {\em Proc. Natl. Acad. Sci. USA}  {\bf 99} (2002) 7267-7274

\bibitem{portercommunitystructurePhA386}  M.A. Porter, P.J. Mucha, M.E.J. Newman, and A.J. Friend,
Physica A  {\bf 386} (2007) 414-438.

\bibitem{communitystructurePorter0902} M.A. Porter, J.P. Onnela, and P.J. Mucha,
Notices of the American Mathematical Society   {\bf 56} (2009) 1082-1097 \& 1164-1166.

\bibitem{implicitcommIJMPC} Xia Meng  Si, Yun Liu and Zhen Jiang  Zhang,
Intern. J. Mod. Phys. C  {\bf 20} (2009) 2013-2026.
 

\bibitem{ballcriticalmass} Ph. Ball, {\it Critical Mass: How One Thing Leads to Another}, Heinemann/Farrar Straus \& Giroux (2004).

\bibitem{reftomedia} This subject of intense interest is  considered and conveniently  approached in   $http://www.talkorigins.org/origins/faqs-creationists.html$

\bibitem{garcia} A. Garcia Cant\`u Ross and M. Ausloos, {\em Scientometrics} {\bf 80}   (2009) 457-472.

 \bibitem{hurd} $http://www.arn.org/blogs/index.php/3/2007/10/14/science_-and_-$   $intelligent_-design_-to_-be_-discussed$

\bibitem{miller} $http://www.brownalumnimagazine.com/november/december-2005/$ $the-evolution-of-ken-miller.html$

\bibitem{bosetti} G. Bosetti, {\em La mia fede}, Reset, Marsilio Ed. (2008).


\bibitem{karate} W. W. Zachary,
J. Anthropol. Res. {\bf 33}  (1977) 452-473.

\bibitem{mormons}  H.R. Bernard, P.D. Kilworth, M.J. Evans,  C. McCarty, and G.A. Shelley,
 Ethnology {\bf 27}  (1988) 155-179.

 \bibitem{newman}
M.E.J.  Newman,
{\em Proc. Natl. Acad. Sci. USA} {\bf 98},  404-409
(2001).

     \bibitem{iina2} I. Hellsten, R. Lambiotte, A. Scharnhorst, and M. Ausloos, 
     Scientometrics {\bf 72} (2007) 469-486.

\bibitem{assortativity} M.E.J. Newman, 
Phys. Rev. Lett. {\bf 89}  (2002) 208701.

\bibitem{mixing} M.E.J. Newman,
Phys. Rev. E {\bf 67}   (2003)  026126. 

  \bibitem{gligorausloos} M. Gligor and M. Ausloos,
 Eur. Phys. J. B {\bf 63} (2008) 533-539

\bibitem{galamdebates} S. Galam, {\it Public debates driven by incomplete scientific
data: the cases of evolution theory, global
warming and H1N1 pandemic influenza}, private communication

    \bibitem{Pennock} R.T. Pennock,   
     Ann. Rev. Genomics Hum. Genet. {\bf 4}  (2003) 143-163.

     \bibitem{amaralPNAS} L.A.N. Amaral, A. Scala, M. Barthelemy, and H.E. Stanley
  {\em Proc. Natl. Acad. Sci. USA} {\bf 97}  (2000) 11149-11152.

\bibitem{jbtest}  C.M. Jarque and A.K. Bera, 
 Intern. Stat. Rev. {\bf 55}   (1987) 1-10.

  \bibitem{newman0303516}
M.E.J.  Newman,
{\it  The structure and function of complex networks},
arXiv:cond-mat/0303516 v1

 

\bibitem{citationnetworkredner}  S. Redner,
Eur. Phys. J. B  {\bf 4}   (1998) 131-134.

\bibitem{citationnetworkRadicchi} F.  Radicchi,  S.  Fortunato, and
C. Castellano,
Proc. Natl. Acad. Sci. USA {\bf 105}   (2008) 17268-17272.

\bibitem{MLE01}  J.P. Nolan, {\it Maximum likelihood estimation and diagnostics for stable distributions}, in O. Barndorff-Nielsen, T. Mikosh, and S. Resnick, editors, {\it  L\' evy Processes: Theory and application}  (Birkh\"auser, Boston, 2001).

\bibitem{MLE07}   A.C.C. Shalizi and M. Newman, Power-law distributions in empirical data, E-print, arXiv:0706.1062 (2007).

\bibitem{GPD52}   H.L. Seal, 
J. Institute of Actuaries {\bf 78} (1952) 115--121 .

\bibitem{Fama65}  E.F. Fama,
 J. Business {\bf 38}  (1965) 34-105.

\bibitem{GPD07}   G Rotundo, M Navarra,
 Physica A  {\bf 382}  (2007)   235-246.


\bibitem{pajek} V. Batagelj and A. Mrvar, {\it Pajek, Program for Analysis and
Visualization of Large Networks
Reference Manual},  Fig. 15, p. 52

\bibitem{milo} R. Milo, S. Shen-Orr, S. Itzkovitz, N. Kashtan, D. Chklovskii, and  U. Alon,
Science    {\bf 298}  (2002) 824-827.

\bibitem{AOILA} F. O. Redelico, A.N. Proto, and M. Ausloos,
 Physica A {\bf 388}  (2009) 3527-3535

\bibitem{ebeling}
I. Hellsten, R. Lambiotte, A. Scharnhorst, and M. Ausloos, 
 {\em Scientometrics} {\bf 72} (2007) 469-486.

\bibitem{extra}  G. Tibely and J. Kertesz,
 Physica A {\bf 387}  (2008) 4982-4984.


 \bibitem{PNAS101.04.5287} F. Radicchi, C. Castellano, F. Cecconi, V. Loreto, and D. Parisi,
  {\em Proc. Natl. Acad. Sci. USA}  {\bf 101} (2004) 2658-2663.

 

\bibitem{bernard} H. R. Bernard, P. D. Killworth, C. McCarty, G.A. Shelley, and S.Robinson.
 Social Networks  {\bf 12}  (1990) 179-215.

\bibitem{CHESS} M. Ausloos, {\it On religion and language evolutions seen through mathematical and agent based
models}, Proceedings CHESS 2009 conference, in press.

       \bibitem{categorical}
       see,  e.g., $http://www.oswego.edu/$ $sims$ $rp /stats/variable_-types.htm$;
or
$http://www.ats.ucla.edu/stat/mult_-pkg/whatstat/nominal_-ordinal_-interval.htm$


     \bibitem{CSC} see $http://www.discovery.org/csc/$

     \bibitem{NCSE} see $http://www.ncseweb.org/$
 
\end{thebibliography}
\end{document}